\documentclass[12pt]{iopart}
\usepackage{amssymb}
\usepackage{natbib}
\usepackage{hyperref}

\newcommand{\newblock}{}
\usepackage{xr}
\usepackage{graphicx}
\usepackage{xcolor}
\usepackage{pdfpages}

\newcommand{\red}[1]{{\color{black}#1}}
\newcommand{\blue}[1]{{\color{black}#1}}
\usepackage[pagewise]{lineno}
\begin{document}

\title[Global Climate Model Bias Correction Using Deep Learning]{Global Climate Model Bias Correction Using Deep Learning}

\author{Abhishek Pasula \& Deepak N. Subramani\footnote{Corresponding Author: deepakns@iisc.ac.in}}

\address{Department of Computational and Data Sciences, Indian Institute of Science, Bengaluru 560012, India.\\
Divecha Center for Climate Change, Indian Institute of Science, Bengaluru 560012, India.}
\ead{deepakns@iisc.ac.in}

\begin{abstract}
Climate change affects ocean temperature, salinity, and sea level, which in turn impacts monsoons and ocean productivity. Future projections by Global Climate Models (GCM) based on shared socioeconomic pathways (SSP) from the Coupled Model Intercomparison Project (CMIP) are widely used to understand the effects of climate change. However, CMIP models have a significant bias compared to reanalysis in the Bay of Bengal. For example, there is a 1.5$^{\circ}$C root mean square error (RMSE) in the sea surface temperature (SST) projections of the climate model CNRM-CM6 compared to the Ocean Reanalysis System (ORAS5). We develop a suite of data-driven deep learning models for bias correction of climate model projections and apply it to correct SST projections of the Bay of Bengal. We propose the use of three different deep neural network architectures: a convolutional encoder-decoder UNet, a Bidirectional LSTM, and a ConvLSTM. We used a baseline linear regression model and the Equi-Distant Cumulative Density Function (EDCDF) bias correction method for comparison and evaluation of the impact of the new deep learning models. All bias correction models are trained using pairs of monthly CMIP6 projections and the corresponding month's ORAS5 as input and output. Historical data (1950-2014) and future projection data (2015-2020) of CNRM-CM6 are used for training and validation, including hyperparameter tuning. Testing is performed on future projection data from 2021 to 2024. We found that the UNet architecture that is trained using a climatology-removed CNRM-CM6 projection as input and climatology-removed ORAS5 as output gives the best bias-corrected projections. Our novel deep learning-based method for correcting CNRM-CM6 data has a 15\% reduction in RMSE and a 12\% increase in the pattern correlation coefficient compared to EDCDF.

\end{abstract}

%
%
%
%
%

\section{Introduction}

Climate change poses a growing and ongoing threat to social and economic well-being and has been characterized by the United Nations as one of the defining issues of our time \citep{stocker2014summary}. The ocean plays an important role in the dynamics of the Earth system, but climate change is modifying the physical and chemical properties of the ocean that have direct consequences for marine ecosystems and indirectly affect the land and atmosphere. The tropical sea surface temperature (SST) is particularly sensitive to climate change, with warming trends that disrupt the critical ocean-atmosphere interactions \citep{jochum2005internal, mohan2021long}. Quantifying the anthropogenic impact on climate change is a controversial subject that the climate research community must address rightly to adopt the correct climate policies. The World Climate Research Programme (WCRP) runs the Coupled Model Intercomparison Project (CMIP), whose data is used regularly by the Intergovernmental Panel on Climate Change (IPCC) in its assessment reports. Since the 1990s, the Coupled Model Intercomparison Project (CMIP) has facilitated the simulation of numerous Global Climate Models (GCMs) from a multitude of international institutes to enhance simulation accuracy and acquire precise scientific insights into the Earth's future climate. The GCMs included in the CMIP Phase 6 (CMIP6) have higher resolution compared to the GCMs of previous phases and use shared socioeconomic pathways (SSP) to offer projections for future climate change mitigation and adaptation efforts \citep{eyring2016overview}. The main demerit of GCMs is their potential for error, which can have significant implications for decision-making and policy making, with considerable biases in simulations that require correction methods to reduce model biases for impact studies \citep{cannon2018multivariate, li2020performance}.

Several statistical bias correction methods have been proposed in the literature for correcting GCM projections \citep{volpi2024legacy}. The Equi-Distant Cumulative Distribution Function (EDCDF) method, a quantile-based mapping method for bias correction \citep{piani2010statistical}, is one of the most popular statistical bias correction methods used for the CMIP6 climate models. However, this method is limited to addressing errors in the simulated frequency distributions in each cell of the grid individually and has other issues such as instability and overfitting \citep{li2010bias, franccois2020multivariate}.

In recent years, deep learning has emerged as a prominent research methodology in various fields. This approach has penetrated almost every facet of our lives, offering intelligent solutions to challenges that were previously intractable or difficult to address. Today, machine learning technology has emerged as a powerful tool for solving difficult problems in climate science. A primary advantage of machine learning (ML) techniques is their ability to extract more pertinent information from multiscale datasets, thereby achieving a closer alignment with ground truth. Recent research has shown that machine learning is particularly effective in weather forecasting, predicting long-term climate trends, and filling in gaps in climate elements, with substantial advances being recorded \citep{reichstein2019deep}. Deep neural networks from computer vision and image processing have been utilized effectively for a variety of tasks within the field of Earth system sciences, including weather forecasting \citep{bi2023accurate, rasp2021data, zhang2023skilful}. Recently, deep learning has been used to correct for bias in Tmin, Tmax \citep{wang2022deep}, precipitation \citep{fulton2023bias, huang2023investigating, huang2024unsupervised}, and wind energy \citep{zhang2021future}.

The goal of the present paper is to explore the use of deep neural networks to perform bias correction in global climate models, with a focus on ocean surface properties. Specifically, we answer the following research questions.
\begin{enumerate}
    \item Can bias correction of climate model projections be performed using deep neural operator learning? If so, what is the mathematical basis of this correction?
    \item What architecture is suitable for the task? How do the different architecture choices perform in terms of figures of merit? 
    \item How do deep models compare against a baseline of the linear model or the statistical Equi-Distant Cumulative Density Function (EDCDF) method?
    \item What data pre-processing strategies are essential for successful model development?
\end{enumerate}

We answer all the above questions and establish that the UNet architecture, a fully convolutional encoder-decoder operator network, is the best choice for climate model bias correction. Detailed ablation and architecture choice studies are completed with applications to correcting the sea surface temperature in the Bay of Bengal. 
        
The remainder of the paper is organized as follows. Section~\ref{sec:method} describes the mathematical problem statement, describes the literature survey of bias correction methods, discusses the choice of neural architectures, and describes the data sets. The main UNet model is developed in Section~\ref{sec:unet}, and the other architecture choices including linear models and statistical baselines are provided in Section~\ref{sec: ablation}. Section~\ref{sec:results} documents the comparison of different bias correction methods. Finally, a summary and future directions are given in Section~\ref{sec: conclusion}.

\section{Method Development} \label{sec:method}
\subsection{Mathematical Problem Statement} \label{sec:problemStmt}
Consider the spatio-temporal coordinates $(r,t)$ in a domain $r \in \Omega \times [0,\infty)$ where 
$\Omega \subset \mathbb{R}^d$ is the spatial domain (d=2 for 2D and d=3 for 3D in space). Within this framework, for a 2D spatial domain, we define $U(x,y,t)$ as the climate projection generated by a General Circulation Model (GCM), while $U^\dagger(x,y,t)$ represents the reanalysis data derived from observational measurements.

We can conceptualize a bias-correcting mathematical operator $\mathcal{G}$ such that $U^\dagger = \mathcal{G}(U)$, which transforms the GCM projections into their bias-corrected counterparts. We propose to approximate this operator using a parameterized neural network $\mathcal{N}_\theta$, trained on historical pairs of climate projections and their reanalysis datasets. Under the assumption that reanalysis data constitute the ground truth representation of the Earth system (specifically, oceanic conditions in our current study), the resulting neural operator effectively functions as a bias correction mechanism for climate projections.

\red{The above formulation is an operator learning framework for bias correction. We do not explicitly train the neural operator with a target to correct any specific moments such as mean, variance, or higher-order moments; rather the mean square error is minimized between the neural operator's output and the reanalysis data. This approach forces the operator to learn the distributional properties \citep{lu2021learning,kovachki2023neural,raonic2023convolutional}.}

\blue{
\subsection{Related Work}
Bias correction of GCM projections could be classified into three types: \textit{(i)} statistical methods, \textit{(ii)} dynamical downscaling, and \textit{(iii)} machine learning methods.

        \paragraph{Statistical methods} Statistical bias correction methods can be categorized as
        univariate and multivariate methods.  Univariate methods address individual variables on their own, while multivariate methods handle biases for each variable and also modify the interdependencies among variables in the climate model \citep{piani2010statistical, piani2012two, cannon2015bias, maraun2016bias, volpi2024legacy}. Interpolation is a widely used bias correction method that adjusts mean and variance using projection and reanalysis (or observational) data. Some extensively used univariate interpolation statistical methods are explained below.
        }
        The delta change approach uses observations as a basis that produces future time series with dynamics similar to current conditions \citep{graham2007assessing}. The linear scaling approach adjusts the monthly mean values and offers corrected data with a consistent variability with the observation data \citep{graham2007assessing}. Local Intensity Scaling (LOCI) combines the advantages of linear scaling with a correction of wet day frequencies (precipitation threshold) \citep{schmidli2006downscaling}. Although the linear scaling above accounts for a bias in the mean, it does not allow differences in the variance to be corrected. The power transformation is a nonlinear correction method in an exponential form to correct/adjust variance statistics \citep{leander2007resampling}. A similar exponential variance scaling method adjusts both the variance and the mean in a stepwise manner \citep{chen2011uncertainty}. Another method called distribution mapping corrects the distribution function of the model value so that it matches the data \citep{sennikovs2009statistical}. The most widely used is the quantile mapping technique \citep{piani2010statistical} that adjusts the distributional properties of the model more closely to match those of the historical observations. There are different versions of quantile mapping methods available, of which the Equi-Distant Cumulative Distribution Function (EDCDF) method is the most popular for bias correction of CMIP6 projections \citep{li2010bias}.

        \blue{In multivariate methods, conditional binning, a two-dimensional conditional bias correction approach for temperature and precipitation, was proposed by \citep{piani2012two}. The temperature is first corrected by using quantile mapping. Second, precipitation and temperature couples are assigned to one of several temperature bins. Finally, using quantile mapping, precipitation data inside each temperature bin are bias-adjusted. Another category of multivariate methods is iterative; the core concept is to repeatedly apply univariate quantile mapping and multivariate correction \citep{cannon2016multivariate} across multiple time scales \citep{mehrotra2016multivariate}.}

        \blue{In nested bias correction, the distributional and persistence bias is corrected from fine to progressively longer time scales \citep{johnson2012nesting}. Frequency-based bias correction works in the frequency space and is independent of specific time scales \citep{nguyen2016correcting}. The wavelet-based signal processing bias correction approach was implemented for the GCM future time series and described the discontinuity in the trend from current to future climate projections \citep{kusumastuti2021signal}. Studies have shown that quantile mapping, especially EDCDF, effectively reduces biases in GCM projections \citep{sachindra2012statistical, yang2018bias, mondal2021projected}. In this work, we have used EDCDF as the baseline for evaluating our neural model to correct the CMIP6 projections (Section~\ref{sec: edcdf}). }
      
        \paragraph{Dynamical downscaling} Dynamical downscaling, in contrast to interpolation and statistical downscaling, can generate dynamically consistent high-resolution climate data with a wide range of physical processes and their complex interactions within the Earth system. Traditional dynamical downscaling of future climate involves combining the initial and lateral boundary conditions of a general circulation model (GCM) with a regional climate model \citep{giorgi2009addressing}. Coordinated Regional Climate Downscaling Experiment (CORDEX) is a global initiative of the World Climate Research Program (WCRP) focused on producing high-resolution climate change projections for different regions of the world \citep{giorgi2015regional}.
            
        \paragraph{Machine learning methods} 
        Machine learning (ML) algorithms can make data-driven predictions using supervised learning that involves learning a mapping function between input and output variables. With the growing popularity and advances in machine learning, there has been an increase in the use of ML techniques with environmental data \citep{reichstein2019deep}. These techniques are used to analyze and model complex ocean and atmospheric data, such as temperature, precipitation, salinity, and currents, to improve our understanding of the atmosphere and ocean dynamics and their impact on climate and weather patterns \citep{ahijevych2016probabilistic, boukabara2020outlook}. Furthermore, ML is being used to develop more accurate and efficient model forecasting, which can have important implications for various sectors such as agriculture, energy, transportation, infrastructure planning, fishing, and offshore energy production \citep{reichstein2019deep}. Deep neural networks, specifically convolutional and transformer models, have the potential to tackle spatio-temporal problems \citep{labe2024exploring, bano2024transferability}. Recently, deep learning has been used to correct for bias in Tmin, Tmax \citep{wang2022deep}, precipitation \citep{fulton2023bias}, and wind energy \citep{zhang2021future}. \red{In the Indian region, the near-surface temperature \citep{dutta_et_al_2022} and ISMR \citep{sharma_et_al_2024} were bias corrected using deep learning models. To our knowledge, bias correction of the CMIP6 ocean forecasts for the Bay of Bengal region has not been attempted using deep learning methods} From the results cited above in other areas of Earth system modeling, the scope for machine learning for CMIP6 bias correction appears to be high. In this paper, we develop a suite of deep neural networks to correct errors in CMIP6 projections in the Bay of Bengal and identify the best architecture for this purpose.
        
\subsection{Neural Architecture Choices} 
The selection of an appropriate neural network architecture $\mathcal{N}_\theta$ for bias correction of climate model projections presents a complex design challenge with multiple dimensions to consider. Several candidate architectures offer distinct advantages for handling spatio-temporal data patterns inherent in climate projections. These include Dense Neural Networks (DNNs), Convolutional Neural Networks (CNNs), Recurrent Neural Networks (RNNs), transformer models, and neural operators.

DNNs offer simplicity and straightforward implementation, though traditional DNNs may struggle to capture the spatial dependencies crucial for climate field correction without significant feature engineering. CNNs excel at extracting spatial features through hierarchical pattern recognition, making them particularly suitable for processing gridded climate data where local patterns and spatial relationships are significant. RNNs, particularly architectures such as Long Short-Term Memory (LSTM) networks, specialize in temporal dependencies, potentially capturing seasonal patterns and long-term trends in climate model errors. With appropriate preprocessing strategies, RNNs can also handle spatial sequences by transforming gridded data into sequential representations through row-wise (latitude) or column-wise (longitude) scanning approaches. This versatility is further enhanced in Convolutional LSTM (ConvLSTM) architectures, which integrate convolutional operations within the LSTM framework, simultaneously capturing both spatial and temporal dependencies. Despite these advantages, training RNNs presents significant challenges, including vanishing and exploding gradients over long sequences, computational inefficiency with increasing sequence lengths, and difficulty in parallelizing the training process – all of which can impede effective learning of long-range climate patterns. Transformer models employ attention-based mechanisms that have demonstrated exceptional capabilities in modeling long-range dependencies, which could be valuable for capturing teleconnections and remote influences in climate systems. However, these models present significant implementation challenges for climate applications. Transformers typically require massive amounts of training data to achieve their potential, which can be problematic given the relatively limited historical climate records available (relative to the data available for language technology training). Their computational demands are exceptionally high, with quadratic scaling of memory and computation with respect to sequence length, making them resource-intensive for high-resolution climate fields. In addition, training stability issues often arise with transformers, requiring careful hyperparameter tuning, learning rate scheduling, and gradient clipping to achieve convergence. The positional encoding schemes that traditional transformers use may also need to be adjusted to irregular grids that are sometimes encountered in climate data. Neural operators such as Convolutional Neural Operators, Fourier Neural Operators, Geometric Neural Operators, and other architectures are sophisticated architectures designed to learn mappings between function spaces, enhancing bias correction generalization.

Beyond architecture selection, the input-output configuration presents additional design considerations. Options include direct mapping from raw climate model outputs to corrected fields, inclusion of auxiliary variables as conditioning information, or embedding coordinate information to enhance spatial awareness. The target output could range from complete field reconstruction to anomaly correction or adjustment of statistical parameters. 

The temporal resolution of the modeling approach represents another critical design choice. Given that most CMIP6 Global Climate Models provide monthly outputs, our neural model development naturally aligns with this discrete monthly timescale. This allows for capturing seasonal cycles and interannual variability while maintaining computational feasibility. Correction models could accept a sequence of several months of spatio-temporal projections or a single month as input and output a sequence of months or a single month. In our ablation study, we tried both options and present the results.

The combination of these architectural choices, input-output configurations, and temporal considerations creates a vast design space of potential neural models. Our approach to navigating this space involves systematic experimentation with representative architectures, guided by physical understanding of climate processes and error patterns in the Bay of Bengal region. Specifically, we develop a convolutional encoder-decoder neural operator model (Section~\ref{sec:unet}), bidirectional LSTM model and ConvLSTM model (Section~\ref{sec: ablation}). We also use simple baseline models, such as linear regression, to quantify the goodness of our bias correction method.

\subsection{Data Sets} \label{sec: data}

In this work, we selected the CNRM-CM6-1-HR Global Climate Model (GCM), which is a component of the CMIP6 suite, from the National Center for Meteorological Research (CNRM), France. Compared to earlier iterations, this model shows less bias in ocean temperature and salinity \citep{voldoire2019evaluation}, and its predictions for ocean heat uptake closely resemble observed data \citep{kuhlbrodt2023historical}. In addition to historical simulations from 1850 to 2014, CNRM-CM6 also provides future projections up to 2100 over four Shared Socioeconomic Pathways (SSPs)—SSP1-2.6, SSP2-4.5, SSP3-7.0, and SSP5-8.5. These scenarios cover various routes for emissions, land usage, and energy use. The CNRM-CM6-1-HR has high spatial resolution with a 25 km ocean grid and a 50 km atmospheric grid, facilitating detailed simulations of atmospheric phenomena, regional flows, and ocean circulation \citep{voldoire2019evaluation}.

The training target for our bias correction model is reanalysis data. We selected Ocean Reanalysis System 5 (ORAS5), developed by the European Center for Medium-Range Weather Forecasts (ECMWF) in this work. ORAS5 is configured with the NEMO ocean model version 3.4.1 \citep{madec2017nemo} at a 25 km horizontal resolution and incorporates multiple observational datasets through a 3D-Var assimilation scheme with a 5-day cycle. The system assimilates sea surface temperature, subsurface temperature, salinity profiles, satellite sea level measurements, and sea ice concentration data. ORAS5 nudges the sea surface salinity to climatology \citep{zuo2019ecmwf}.

Historical CNRM-CM6-1-HR and ORAS5 employ an identical ocean model and span the same modeling period, but differ in initial and forcing conditions. CNRM-CM6-1-HR considers observations about greenhouse gas concentrations, global gridded land use forcing, solar forcing, stratosphere aerosol data set, Atmospheric Model Intercomparison Project (AMIP) SST and sea ice concentration, ozone chemistry and aerosol forcing \citep{voldoire2019evaluation}. ORAS5 uses initial conditions from ERA-40 \citep{uppala2005era}, ERA-Interim forcing fields \citep{dee2011era}, and different ocean observations assimilated in an operational ensemble reanalysis \citep{zuo2019ecmwf}. Furthermore, CNRM-CM6 and ERA5 use different atmosphere models. ERA5 uses the Integrated Forecast System (IFS) model coupled with a land surface model (HTESSEL), while CNRM-CM6 uses the atmosphere-ocean general circulation model (AOGCM) with different initial conditions. ORAS5 includes freshwater discharge, while CNRM uses total runoff integration pathways to model river discharge \citep{seferian2019evaluation}.

\subsection{Training, Validation and Testing}
        
CNRM-CM6 historical data span from 1958 to 2014, and projections extend from 2015 to 2100. For training and validation, we utilized data from 1958 to 2020 (972 months), dividing it into 768 months for training and 204 for validation. The test data set comprises data from 2021 to June 2024. 

The monthly climatology is derived from the ORAS5 reanalysis and subtracted from the input and target data. Our ablation study indicates that this climatology removal produces optimal results.
\blue{Such pre-processing is widely used in the GCM bias correction methods \citep{tabor2010globally, kim2014improvement, navarro2020high, xu2021bias}.}
All monthly climatology-removed data from CNRM-CM6 and ORAS5 were resized to $128 \times 128$ for our UNet model. This choice is to ensure that the dimensions have a power of 2 and that the original data resolution is not lost \blue{due to the symmetric encoder-decoder structure and skip connections of the UNet.} ORAS5 is used as the target for all SSP scenarios. 


\section{UNet: Fully Convolutional Encoder-Decoder Architecture}\label{sec:unet}

\begin{figure}[htbp]
            \centering
            \includegraphics[width=\linewidth]{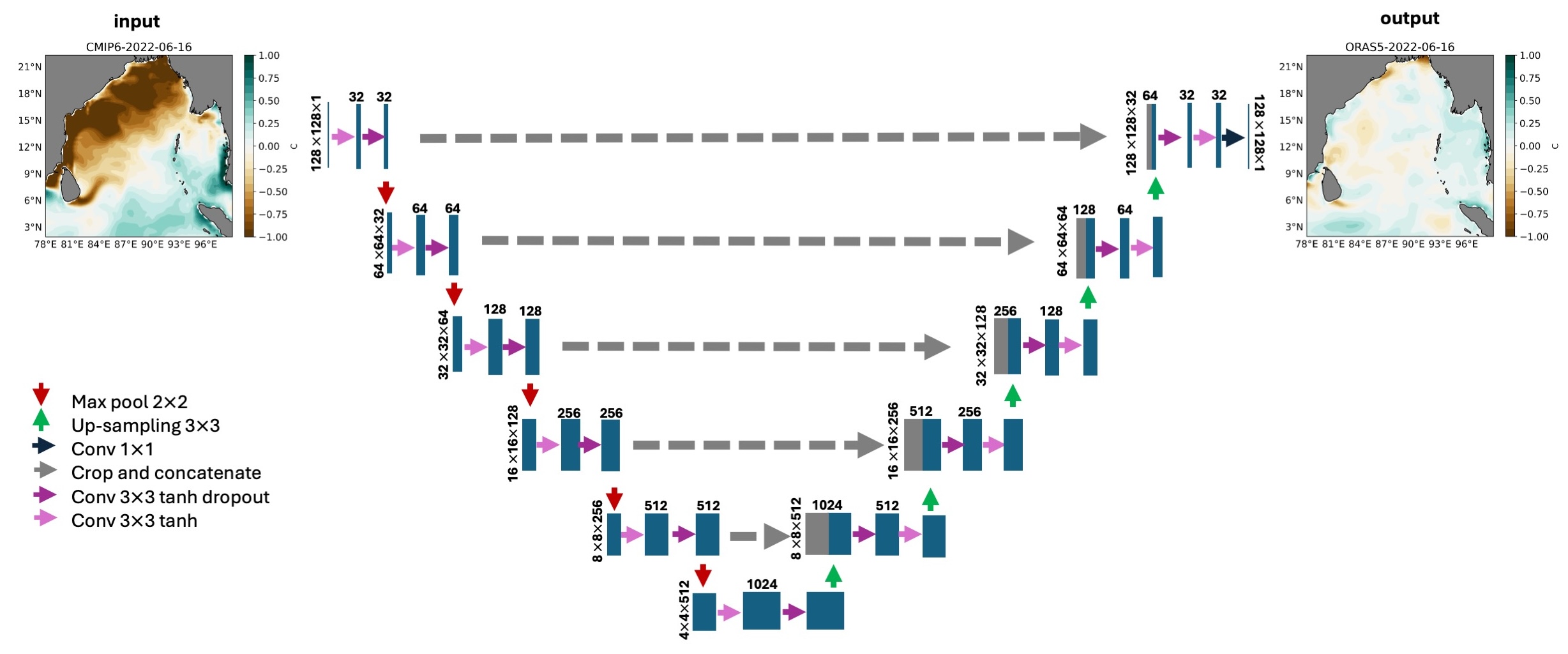}
            \caption{Schematic of the UNet fully convolutional encoder-decoder architecture for bias correction of CNRM-CM6.}\label{fig: unetdes}
        \end{figure}

        Figure~\ref{fig: unetdes} shows the UNet neural network architecture that we developed for bias correction of the climate model projections. This architecture was selected to combine the global context with local high-resolution features, making it particularly effective for our application, where fine details are as important as the overall structure. 
        
        The input to the neural model is the pre-processed climate model projection, and the output is the bias corrected projection. During training, the target is supplied with the ORAS5 fields corresponding to the input climate model projections.

        The UNet architecture takes as input a $128 \times 128$ time slice of the monthly projection by CNRM-CM6. The CNRM-CM6 and ORAS5 data are available with a resolution of 85$\times$85 in the Bay of Bengal study domain. These fields are bilinearly interpolated to $128 \times 128$. The processing is done by using a fully convolutional encoder-decoder structure. The encoder path comprises five downsampling stages using $3 \times 3$ convolutional layers with tanh activation and dropout, followed by $2 \times 2$ max pooling, with the number of filters doubling at each stage (32, 64, 128, 256, 512). The bottleneck contains two $3 \times 3$ convolutions with 1024 filters. The decoder mirrors this structure with five upsampling stages using $2 \times 2$ transposed convolutions, where each stage combines the upsampled features with the corresponding encoder features through skip connections (concatenate operation). Each decoder block includes two $3 \times 3$ transpose convolutional layers with tanh activation, and the final output is produced by a $1 \times 1$ convolution. The exact choice of the number of filters at each stage was reached during hyperparameter tuning on the validation data. We implement Python-based data processing tools and a UNet neural network implementation for bias correction of CNRM-CM6 SST and make our code available on GitHub (https://github.com/AbhiPasula/Climate-Model-SST-Bias-Correction-Toolkit). 

        \red{The above convolutional neural operator \citep{raonic2023convolutional} uses convolutional layers to account for spatial dependencies in bias correction, thereby enhancing the overall accuracy of the deep model.} 

        \subsection{Data Pre-processing}
        Through extensive experimentation, we found that preprocessing the input climate projections and output ORAS5 reanalysis using the monthly climatology from ORAS5 achieves the best bias correction results. Using ORAS5 data from 1958 to 2014, we calculate the mean of each month. Then, this calculated mean is removed from the climate model projections for each month and the corresponding ORAS5 target. During inference, the monthly climatology is added back to the output of our UNet model. We hypothesize that using climatology-removed data for training and inference allows the neural model to better learn the distribution of data.
        
        \subsection{Hyperparameters}
       The initial weights are sampled from the He normal distribution. The Adam optimizer is utilized for training. The validation set, comprising 20\% of data from 1958 to 2020, includes historical simulations and projections. Hyperparameters are tuned through cross-validation on this validation set, selecting the model with the best performance for final use. Sweeping through batch sizes from 32 to 128 and testing various learning rates, we found that a batch size of 64 and a learning rate of 0.0001 are ideal. The architecture uses the tanh activation function in the convolutional layers to better handle the ocean data values, including positive and negative anomalies. Dropout layers (0.2) are incorporated between the encoder and decoder blocks to prevent overfitting, which is especially important given the spatial and temporal correlations in the climate projections. 
   
\section{Other Architecture Choices and Baselines} \label{sec: ablation}
        To arrive at the final UNet model, we performed several ablation studies and also experimented with other architectural choices. We trained two variants of sequence-to-sequence deep learning models based on recurrent neural architectures. First, we flattened the ocean grid points in a row-major (latitude) and column-major (longitude) form and used a bidirectional LSTM model. In this choice, we used the projection of one month as input, and the corrected projection of the same month as the output was obtained. In another sequence-to-sequence model, we used a ConvLSTM spatio-temporal model that accepted a sequence of 4 month projections as input and produced a 1 month corrected projection as the output. For baseline, we developed a linear regression model that operates on a flattened vector of ocean grid points. We also used the statistical EDCDF method as a baseline to compare the performance of all machine learning models. In the UNet ablation study, four UNet architectural variants were systematically evaluated to assess the contribution of different design components to model performance. The supporting information provides a comprehensive description of further ablation studies for the UNet architecture. Section S2 contains a thorough analysis of the UNet ablation study, while Section S7 presents findings from an additional ablation study that uses HadGem3 data as input to the UNet model trained on CNRM-CM6 data.

        \subsection{BiLSTM} \label{sec: bilstm_arch}
\begin{figure}[htbp]
                \centering
                \includegraphics{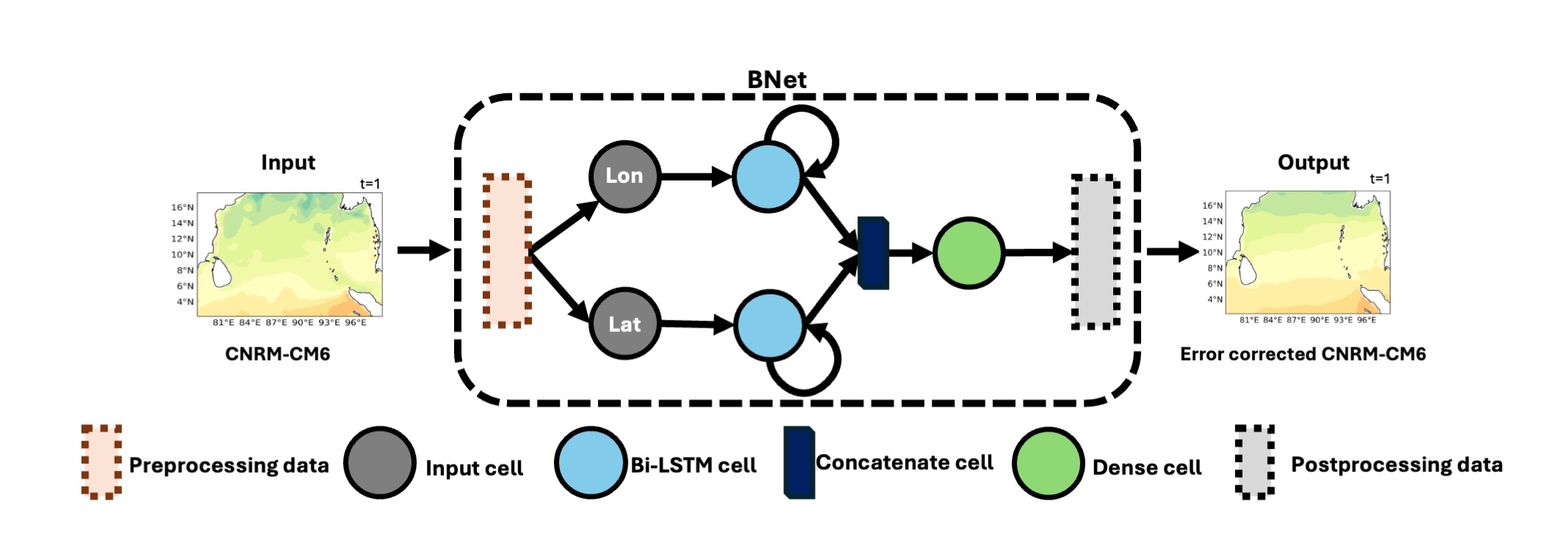}
                \caption{
                Schematic of the BiLSTM architecture for bias correction of CNRM-CM6.
                }\label{fig: bilstm_arch}
            \end{figure}

            The BiLSTM model takes monthly CNRM-CM6 projections as input and gives a corrected field as the output. During training, ORAS5 is used as the ground truth target. The BiLSTM model treats ocean variables in a row-major and column-major format and applies 1D convolutional layers to extract their features. A BiLSTM layer with these encoded spatial representations is then used to estimate the bias-corrected fields. Figure~\ref{fig: bilstm_arch} shows a schematic of the architecture. We performed ablations with only row-major and column-major data pre-processing and found that using both as two channels in the BiLSTM architecture is the best model. 
            
        \subsection{ConvLSTM} \label{sec: convlstm_arch}

\begin{figure}[htbp]
                \centering
                \includegraphics[width=0.95\columnwidth]{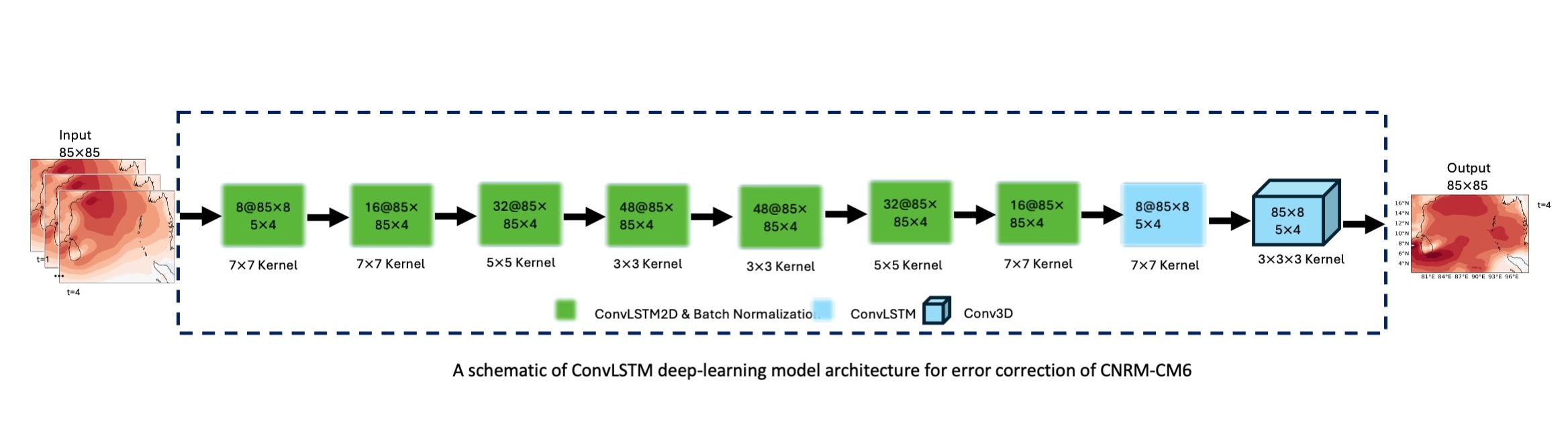}
                \caption{Schematic of the ConvLSTM architecture for bias correction of CNRM-CM6.
                }\label{fig: convlstm_arch}
            \end{figure}

            The ConvLSTM model is a spatio-temporal model that applies convolutions on spatial 2D data to obtain representations on which LSTM architecture is applied pixel-by-pixel. This model can be used to accept any sequence length as input and output. During training, the CNRM-CM6 and ORAS5 data must be prepared appropriately. We performed ablations with multiple input and output sequence lengths. We found that the ConvLSTM model takes 4 month input sequences, each with dimensions of $85\times 85$ pixels, and a 1 month output was the best. Figure~\ref{fig: convlstm_arch} illustrates the architecture scheme of the ConvLSTM model. The ConvLSTM layers begin by enhancing the feature representation from 8 filters using a $7 \times 7$ kernel to 48 filters with a $3 \times 3$ kernel, before symmetrically reducing back to 8 filters. The final layer uses a Conv3D operation with a $3\times 3\times 3$ kernel to reconstruct the output sequence. 

    \subsection{Baseline Models}
    \paragraph{Linear Regression}
        As a simple baseline, we develop a linear regression model. We implemented a pixel-wise linear regression, training for each individual pixel of the ocean field. Linear regression takes the flattened vector of the ocean field as input and predicts the corrected field for that specific pixel location.

    \paragraph{EDCDF Statistical Method} \label{sec: edcdf}
        The Equi Distant Cumulative Density Function (EDCDF) method is a statistical bias correction method popularly used for correcting climate model projections. EDCDF corrects the projections of the model by comparing the output of the model with observations in the historical period \citep{li2010bias}. This allows for bias correction of the CMIP6 fields by using the resulting CDFs to correct discrepancies between the model and observations. The bias-corrected CMIP6 field can be expressed as follows.
            \begin{equation}
                x_{m-p\_adjusted} =  x_{m-p}+F^{-1}_{o-c}[F_{m-p}(x_{m-p})]-F^{-1}_{m-c}[F_{m-p}(x_{m-p})]\,,
            \end{equation}
            where $x_{m-p\_adjusted}$ represents the bias-corrected results, $x_{m-p}$ denotes the raw model projections,
            $F_{m-p}$ is the CDF of the CMIP6 model simulations $m$ in the projection period $p$, $F^{-1}_{o-c}$ and $F^{-1}_{m-c}$ stands for the corresponding quantile functions for observations $o$ and model simulations $m$ in the historical training period $c$, respectively.
        
        The EDCDF method assumes that the statistical relationship between observations and GCM projections during the training period remains valid for future projections at a given percentile. Its application to the CNRM-CM6 SST correction results in relatively higher RMSE, as shown in the results.
        
    \section{Results and Analysis}\label{sec:results}
            We trained our UNet architecture and all other architecture choices and present significant results here. First, we compare the RMSE for the test years of SST in the Bay of Bengal in Section~\ref{sec: comparison_test}, followed by a long-term analysis of the trend of SST in Section~\ref{sec: comparison_future}. Next, we show the monthly projections of CNRM-CM6 and the corresponding corrected projections in Section~\ref{sec: monthly}. 
    \subsection{Comparison of models in test period}\label{sec: comparison_test}

        \red{
        The RMSE and pattern correlation coefficient (PCC) of SST projections in the Bay of Bengal for the test years are shown in Table \ref{tab: rmse_ablation_study} and Table \ref{tab: pcc_ablation_study} respectively.
         The raw CNRM-CM6 climate model shows the highest RMSE values (1.22-1.34) and lowest PCC values (0.57-0.58), indicating significant bias and weak correlation with ORAS5 reanalysis. Among the correction methods, UNet consistently shows superior performance across both metrics, achieving the lowest RMSE values ranging from 0.4920 for SSP5 to 0.5018 for SSP1, while simultaneously maintaining the highest PCC values (0.7613-0.7768), demonstrating both superior accuracy and strong correlation with ORAS5. Linear regression shows the weakest performance among correction methods with considerably higher RMSE values (0.77-0.82) and lower PCC values (0.6739-0.6990), highlighting the limitations of simple statistical approaches for capturing complex climate patterns. EDCDF achieves moderate performance with RMSE values of 0.62-0.79 and PCC values of 0.6515-0.6871, while BiLSTM and ConvLSTM demonstrate comparable but inferior performance to UNet, with RMSE values around 0.53-0.59 and PCC values of 0.6817-0.7151. These quantitative results provide strong evidence that UNet offers the most reliable approach for correcting climate model errors in the Bay of Bengal region, combining both the highest accuracy (lowest RMSE) and strongest correlation (highest PCC) with observed SST.
        }
        
        \begin{table}
        \caption{RMSE in SST projections of test years (2021-2024) obtained by different CNRM-CM6 correction methods. The best values are highlighted in bold text.}
        \label{tab: rmse_ablation_study}

        \begin{tabular}{|c|c|c|c|c|c|c|}

\hline
\textbf{Scenario} & \textbf{CNRM} & \begin{tabular}[c]{@{}c@{}} \textbf{Linear} \\ \textbf{Regression}\end{tabular} & \textbf{EDCDF} & \textbf{UNet} & \textbf{ConvLSTM} & \textbf{BiLSTM} \\
\hline \hline
SSP1              & 1.2264   &0.7723       & 0.6871          & \textbf{0.5018}        & 0.5830              & 0.5210          \\
\hline

SSP2              & 1.3480 & 0. 7483       & 0.6352         & \textbf{0.5091}        & 0.5573            & 0.5340          \\
\hline

SSP3              & 1.3494  &0.8053       & 0.6255         & \textbf{0.5092}        & 0.6968            & 0.5261          \\
\hline

SSP5              & 1.2368  &0.8253       & 0.7898         & \textbf{0.4920}        & 0.5906            & 0.5046 \\
\hline

\end{tabular}
\end{table}

\begin{table}

        \caption{PCC in SST projections of test years (2021-2024) obtained by different CNRM-CM6 correction methods. The best values are highlighted in bold text.}
        \label{tab: pcc_ablation_study}

        \begin{tabular}{|c|c|c|c|c|c|c|}

\hline
\textbf{Scenario} & \textbf{CNRM} & \begin{tabular}[c]{@{}c@{}} \textbf{Linear} \\ \textbf{Regression}\end{tabular} & \textbf{EDCDF} & \textbf{UNet} & \textbf{ConvLSTM} & \textbf{BiLSTM} \\
\hline \hline
SSP1              & 0.5747   &0.6990       & 0.6812          & \textbf{0.7768}        & 0.6845              & 0.6941          \\
\hline

SSP2              & 0.5721 & 0.6897       & 0.6750         & \textbf{0.7613}        & 0.7151            & 0.7049          \\
\hline

SSP3              & 0.5743  & 0.6866       & 0.6809         & \textbf{0.7615}        & 0.7058            & 0.6952          \\
\hline

SSP5              & 0.7285  & 0.6739       & 0.6515         & \textbf{0.7666}        & 0.6992            & 0.6817 \\
\hline

\end{tabular}
\end{table}

        \subsection{Comparison of models in future projections}\label{sec: comparison_future}
\begin{figure}[htbp]
            \centering
            \includegraphics[width=\columnwidth]{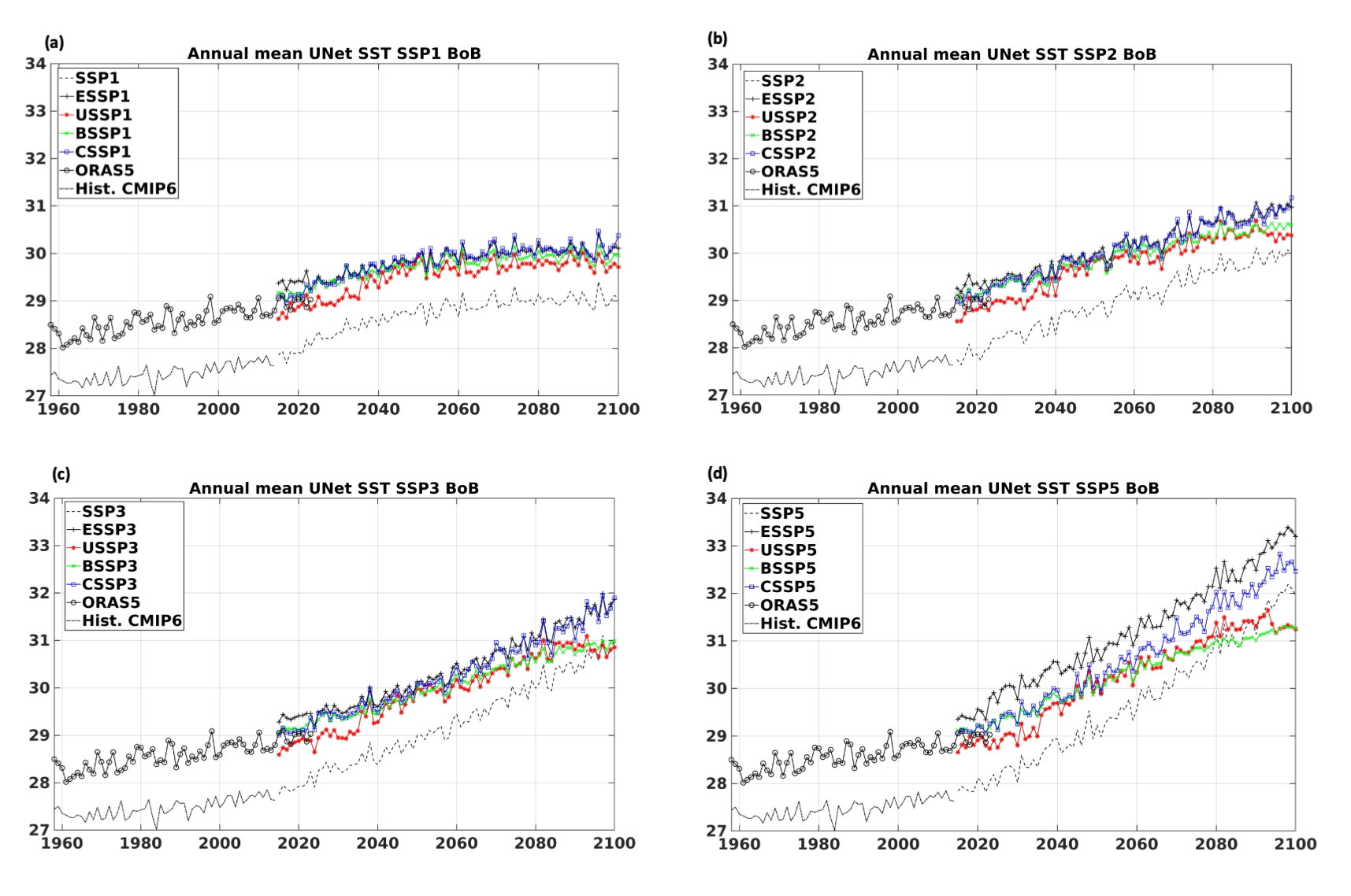}
            \caption{Annual mean SST from CNRM-CM6 projections, corrected projections from EDCDF, UNet, BiLSTM, and ConvLSTM.
            }\label{fig: ablation}
        \end{figure}

        The long term annual sea surface temperature (SST) projections for the Bay of Bengal from 1958 to 2100 for four different SSPs, SSP1 (SSP1-2.6), SSP2 (SSP2-4.5), SSP3 (SSP3-7.0), and SSP5 (SSP5-8.5), are shown in Fig.~\ref{fig: ablation}. Historical CNRM-CM6 (CMIP6) consistently underestimates the temperatures compared to ORAS5 reanalysis. 
        \blue{ A noticeable break in trend is evident when comparing historical to future climate projections. This artifact is expected for statistically bias-corrected time series projections \citep{kusumastuti2021signal}.}
        The long-term trend of SST shows significant variations between different correction methods and SSP scenarios. SST projections indicate substantial warming, with the UNet-corrected SSP5 scenario projecting increases of 3-4$^{\circ}$C by 2100, while the more conservative SSP1 scenario suggests modest warming of 1-1.5$^{\circ}$C. Despite their varying intensities, the consistent upward trend across all scenarios strongly suggests continued ocean warming through the twentieth century, with the magnitude of warming heavily dependent on future greenhouse gas emissions and climate policy decisions. Comparison between SSP scenarios reveals distinct warming trajectories that reflect their underlying socioeconomic and emission assumptions. SSP5-8.5 demonstrates the most dramatic warming, with temperatures reaching approximately 33$^o$C by 2100 (Fig. \ref{fig: ablation}(d)), SSP3-7.0 follows a similar but slightly moderated pattern, with end-century temperatures around 32$^o$C (Fig. \ref{fig: ablation}(c)), SSP2-4.5 shows an intermediate warming path with temperatures approaching 31$^o$C by 2100 (Fig. \ref{fig: ablation}(b)), SSP1-2.6, representing stringent mitigation measures, shows the least warming with temperatures stabilizing around 30$^o$C after 2060 (Fig. \ref{fig: ablation}(a)). The magnitude of warming indicated by the UNet-corrected projections is less severe than that indicated by the EDCDF-corrected SST projections. \blue{The EDCDF correction method, as it is based on quantiles, often inflates trends, resulting in significant deviation in the long-term SST trend in the Bay of Bengal \citep{cannon2015bias, maraun2016bias}.}     
        The BiLSTM correction does not effectively capture the variability in the annual mean projections, producing overly smoothed temperature trends. ConvLSTM projections align more closely with EDCDF-corrected projections. UNet captures both the magnitude and variability of SST changes in the GCM bias correction, suggesting that future warming in the Bay of Bengal may differ from previous projections, with implications for regional climate patterns, monsoon behavior, and marine ecosystems. Overall, we see that the UNet correction has the lowest RMSE and captures the variability accurately. 

        \red{Our comparison results show that for the task of bias correction, the UNet architecture,  a simple convolutional neural operator \citep{raonic2023convolutional}, is ideal. Other sequential processing models such as ConvLSTM and the Bi-LSTM models are sub-optimal for this application. In this study, we restricted our focus to convolutional and recurrent architectures and did not use transformer models \citep{liu2021swin,khan2022transformers,han2022survey}. This choice is primarily due to computational demands and limited data availability for training and validation, which are insufficient to train large transformer models. For example, the Swin transformer has a much higher parameter count than our UNet \citep{xiang2022spatiotemporal, zhong2024investigating}. However, in the future, these heavy models could be implemented for the bias correction problem and compared against the convolutional neural operator used in this paper.}

    \subsection{Analysis of UNet Corrected CNRM-CM6 SSP2-4.5 SST Projections in 2021} \label{sec: monthly}

\begin{figure}[htbp]
            \centering
            \includegraphics[width=1\linewidth]{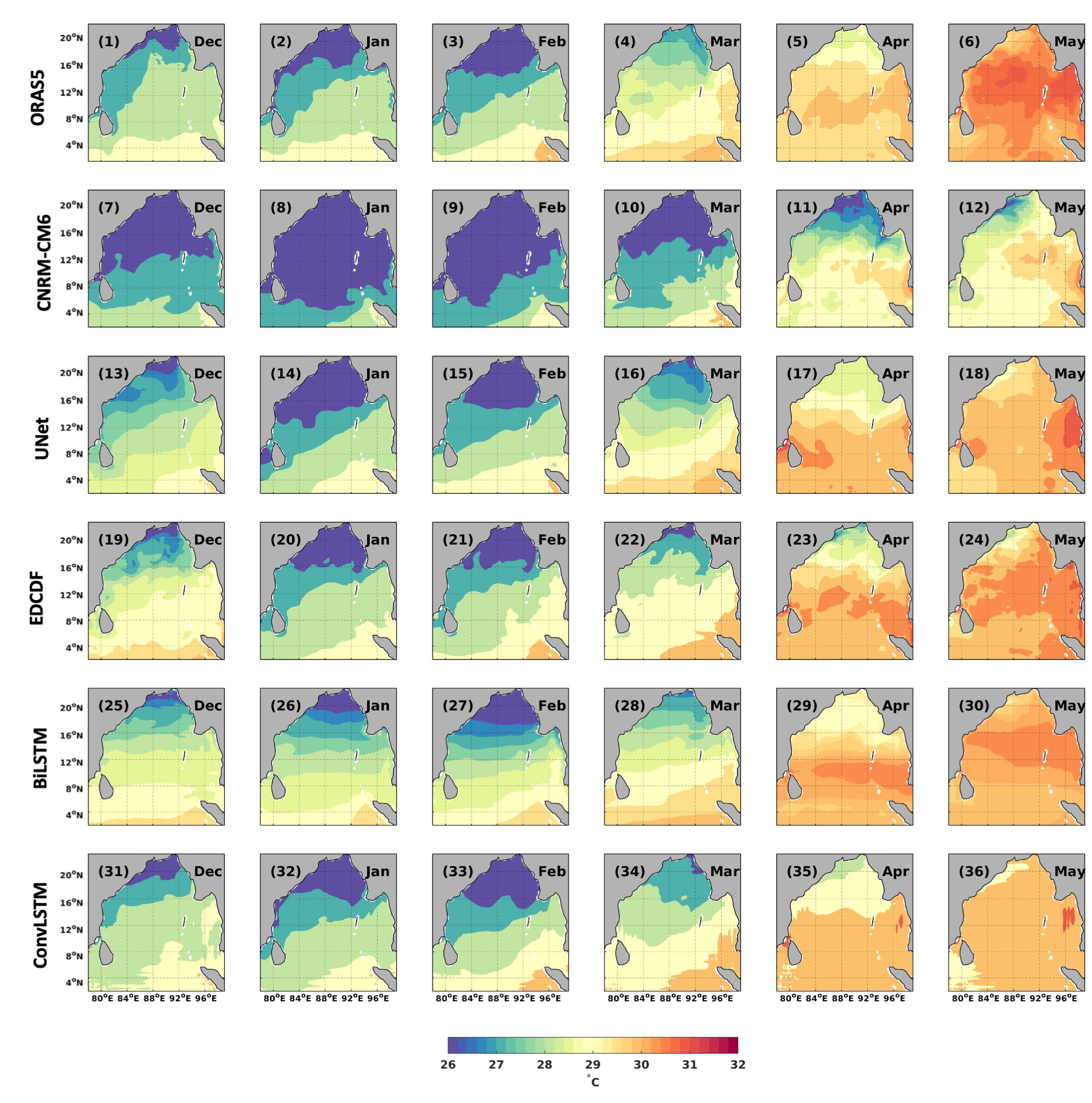}
            \caption{Monthly sea surface temperature (SST) in the Bay of Bengal during 2021 Dec to May, comparing ORAS5 reanalysis data, raw CNRM-CM6 model projections (CNRM-CM6), UNet-corrected SST (UNet), EDCDF corrected SST (EDCDF), BiLSTM corrected SST (BiLSTM), ConvLSTM corrected SST (ConvLSTM).}\label{fig: sst_monthly_2021_1}        
        \end{figure}

\begin{figure}[htbp]
            \centering
            \includegraphics[width=1\linewidth]{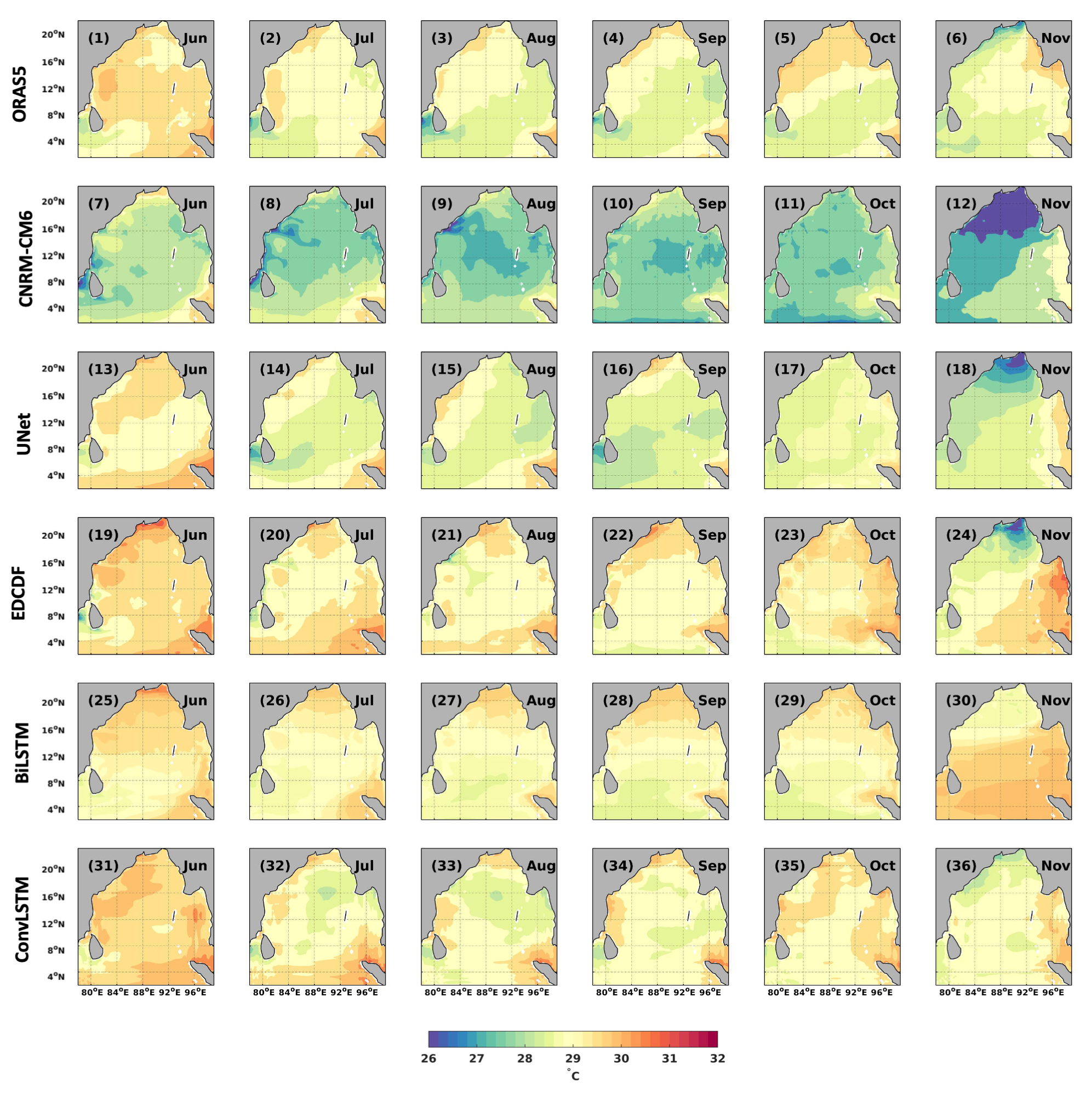}
            \caption{Monthly sea surface temperature (SST) in the Bay of Bengal during 2021 June to Nov, comparing ORAS5 reanalysis data, raw CNRM-CM6 model projections (CNRM-CM6), UNet-corrected SST (UNet), EDCDF corrected SST (EDCDF), BiLSTM corrected SST (BiLSTM), ConvLSTM corrected SST (ConvLSTM).}\label{fig: sst_monthly_2021_2}        
        \end{figure}

        Figures~\ref{fig: sst_monthly_2021_1} and ~\ref{fig: sst_monthly_2021_2} display the monthly SST for 2021 from the reanalysis (ORAS5), raw projections CNRM-CM6 SSP2-4.5 (CNRM-CM6), projections corrected for UNet (UNet), EDCDF corrected SST (EDCDF), BiLSTM corrected SST (BiLSTM) and ConvLSTM corrected SST (ConvLSTM) in the Bay of Bengal. Henceforth, we discuss only the SSP2-4.5 scenario and present the analysis for other SSP scenarios in the supplementary material. 

        \paragraph{Winter} The winter period (December-February) exhibits a pronounced north-south temperature gradient in the Bay of Bengal. The ORAS5 reanalysis shows temperatures of 26-27$^o$C in the northern bay, gradually increasing to 28-29$^o$C in the southern regions. CNRM-CM6 shows a significant cold bias during this period, with temperatures 1-2$^o$C lower than in the reanalysis, particularly in the northern and central regions where temperatures fall below 26$^o$C. In contrast, the UNet corrections substantially improve these patterns, closely matching the ORAS5 reanalysis in both spatial distribution and temperature values. EDCDF correction tends to overestimate temperatures in the northern bay, particularly in December and January, failing to accurately capture the gradient intensity. BiLSTM performs similarly to EDCDF, with corrections showing excessive warming in the northern regions while maintaining reasonable patterns in the south. ConvLSTM correction reproduces the general gradient, but misses the finer circulation features, particularly along the eastern boundary and in regions influenced by the WMC.  

        \paragraph{Pre-monsoon} The pre-monsoon season (March-May) demonstrates progressive warming across the basin. March initiates this warming phase, with temperatures beginning to increase, particularly in the eastern bay. April shows substantial warming and the ORAS5 reanalysis shows temperatures reaching 30$^o$C in the central and eastern regions, marking the development of the warm pool. May represents peak pre-monsoon conditions with temperatures exceeding 31$^o$C in parts of the central and northern bay. Throughout this period, CNRM-CM6 consistently underestimates temperatures by 1-2$^o$C and fails to adequately capture the formation of this warm pool. UNet corrections significantly improve these representations, accurately depicting both the spatial extent and intensity of the warming patterns, particularly in the critical central and eastern regions. EDCDF overestimates temperatures in the northern and western regions in March but shows improved patterns in April and May. BiLSTM correction generates a more uniform temperature field, albeit at the expense of mesoscale features in the eastern bay. In contrast, the ConvLSTM correction demonstrates the least effective performance among correction techniques, failing to capture essential warming patterns in the central bay.

        \paragraph{Monsoon} The monsoon season (June-September) shows distinctive SST patterns associated with monsoon circulation. June marks the onset of the monsoon, with ORAS5 showing temperatures of 29-30$^o$C in most of the bay and slightly cooler waters in regions affected by the monsoon. July and August demonstrate the established monsoon pattern with the Summer Monsoon Current (SMC) evident south of Sri Lanka and the Western Boundary Current (WBC) flowing northward along the Indian coast. September shows the beginning of monsoon withdrawal with slight warming in parts of the northern bay. CNRM-CM6 exhibits a cold bias throughout this period, particularly in the central bay, and fails to accurately represent the SMC's thermal signature. UNet corrections substantially improve these representations, closely matching ORAS5's depiction of basin-wide patterns and localized features associated with monsoon circulation. EDCDF correction shows slightly cooler temperatures than observed, particularly in the central bay. BiLSTM correction captures the general spatial pattern, but produces a more homogenized structure. ConvLSTM correction generates temperature patterns closer to CNRM-CM6 in August and September, particularly along the western boundary, failing to capture the influence of the EICC.

        \paragraph{Post-monsoon} The post-monsoon period (October-November) transitions to winter conditions. October maintains relatively warm temperatures, with ORAS5 showing values around 29- 30$^o$C across much of the bay, while November displays the beginning of the winter cooling pattern with decreasing temperatures in the northern regions and the establishment of a north-south gradient. The East India Coastal Current (EICC) begins its seasonal reversal to southward flow during this period, influencing temperature patterns along the western boundary. CNRM-CM6 shows its most significant cold bias during this transition period, particularly in November, where the temperatures in the northern bay are underestimated by more than 2$^o$C. UNet corrections demonstrate remarkable skill in reproducing the observed ORAS5 patterns, accurately capturing both the spatial distribution of temperatures and the onset of winter cooling, while accurately representing the thermal influence of the EICC along the western boundary. EDCDF correction shows excessive cooling in the northern bay in November. BiLSTM correction captures the general transition pattern, but does not capture important features, particularly along the western boundary. ConvLSTM correction fails to capture the development of the north-south gradient in November and misses the thermal signature of the EICC.

        The difference plots of each correction method on SSP2 with ORAS5 are shown in the supplementary information in section S3. Furthermore, similar analyses of the corrected projections of SSP1, SSP3, and SSP5 are also shown in the supplementary information sections S2, S4, and S5.

\red{
        \subsection{Limitations} \label{sec:limitations}

        The present UNet model (and other architectures) work well with the particular climate model used here, namely, CNRM-CM6. However, when another climate model output is used for bias correction with the model trained on CNRM-CM6, the performance is not as good as a dedicated model trained for that climate model. A study using UNet trained with the CNRM-CM6 forecast, when applied with HadGEM3 \citep{andrews2020historical}, shows few discrepancies compared to ORAS5, particularly in the transition months between seasons and extreme warming. A possibility for future research is to attempt fine-tuning or transfer learning from a bias corrector trained on one climate model and use it for another. A different approach is to employ multiple climate models during the training itself to leverage the generalizability of neural operators.   
}
    \section{Conclusion} \label{sec: conclusion}

        Our study demonstrates that deep neural operator learning can effectively correct biases in the climate projections by global climate models in the Bay of Bengal. We established that the U-Net architecture with an encoder-decoder structure and skip connections provides the mathematical foundation necessary to capture both spatial and temporal patterns in climate data. This approach significantly outperforms traditional methods and was trained on climatology-removed CNRM-CM6 inputs to focus on anomaly patterns and predict the corresponding corrected anomalies. Our experiments revealed that pre-processing the data by removing climatology is essential to obtain state-of-the-art correction results. 
        
        As a baseline, linear regression models and statistical EDCDF methods were also evaluated. We found that baseline models provide basic correction capabilities but fail to capture the complex nonlinear relationships present in climate projections. Although ConvLSTM produced reasonable annual mean projections aligned with statistical methods, only the U-Net model consistently captured both spatial and temporal variability, outperforming all alternatives in reducing errors in the climate projections. This confirmed U-Net as the most suitable architecture for GCM error correction in the Bay of Bengal. Performance evaluation demonstrated a 15\% reduction in RMSE and a 12\% increase in PCC in SST projections compared to the widely used Equidistant Cumulative Distribution Function (EDCDF) method. This improvement highlights the ability of deep learning approaches to model complex, nonlinear, and spatially dependent bias structures inherent in global climate models.

        Looking ahead, the proposed framework offers promising opportunities for application in other ocean basins and climate variables, including ocean currents and atmospheric parameters. The bias correction capability of UNet could be extended to other dynamical variables such as Dynamic Sea Level and Salinity. The oceanographic implications of these corrected projections have also been explored in other papers. In the future, similar architecture development could be completed for atmospheric variables to get a complete picture of the future climate scenarios. Our work presents a structured approach and template for such future development.

\section*{Data Availability}
All data used in this work are available for download from the respective data providers, with links provided in the manuscript. Our code and data produced from this work are available on GitHub (https://github.com/AbhiPasula/Climate-Model-SST-Bias-Correction-Toolkit). 

   \section*{Acknowledgment}
 The authors thank the Ministry of Earth Sciences for the grant that allowed the purchase of the compute resources used in model training. The authors thank the members of the QUEST Lab at IISc for insightful discussion.

 \section*{Statements and Declarations}
\subsection*{Funding Information}
This work was supported by the Ministry of Earth Sciences (grant number MoES/36/OOIS/Extra/84/2022).
\subsection*{Competing Interests}
The authors have no relevant financial or non-financial interests to disclose.
\bibliographystyle{elsarticle-harv}


\newpage


\section*{Supplementary material}

# Supporting Information for Global Climate Model Bias Correction using Data Driven Deep Learning

Abhishek Pasula, Deepak N. Subramani

This document presents additional analyses supporting the Global Climate Model Bias Correction using Data Driven Deep Learning.

## S1 UNet Training

The training and validation loss of the UNet model for GCM SST correction is shown in Fig. S1. The blue line represents the training loss (labeled "train sst"), which decreases rapidly in the first 100 epochs and gradually declines, eventually approaching zero by the end of training. The orange line shows the validation loss (labeled "val loss sst"), initially exhibiting more fluctuation while following a general downward trend. The validation loss stabilizes around 0.2. The best model is selected based on the minimum validation loss, ensuring optimal generalization performance.
.

## S2 Analysis of UNet Corrected CNRM-CM6 SSP1-2.6 SST Projections in 2021

Figure S2 and S3 display the monthly SST for 2021 from reanalysis (ORAS5), raw CNRM-CM6 SSP1-2.6 projections (CNRM-CM6), UNet-corrected projections (UNet), EDCDF corrected SST (EDCDF), BiLSTM corrected SST (BiLSTM), and ConvLSTM corrected SST (ConvLSTM) in the Bay of Bengal.

**Winter** During winter months under the SSP1 scenario, the Bay of Bengal exhibits a well-defined north-south temperature gradient with characteristic cooler waters in the northern regions. ORAS5 reanalysis show temperatures ranging from 26-27$^o$C in the northern bay, gradually increasing to 28-29$^o$C in southern regions. The CNRM-CM6 model demonstrates a consistent cold bias during this period, particularly in the northern bay where temperatures

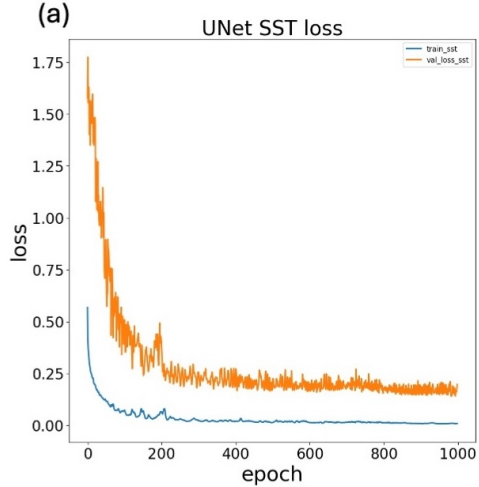

Figure S1: The training and validation loss for the UNet model of (a) SST, and (b) DSL.

are underestimated by 1-2$^oC$ compared to ORAS5. The spatial extent of cooler waters (below 26$^oC$) is exaggerated in the raw CNRM-CM6 output, extending too far southward. Among correction techniques, UNet shows the best performance by accurately reproducing both the temperature values and spatial distribution patterns of the ORAS5 reanalysis. The UNet corrections effectively capture the gradual north-south gradient and maintain appropriate temperature boundaries. EDCDF shows a tendency toward temperature overestimation in the northern bay during December and January, failing to fully represent the intensity of the gradient. BiLSTM exhibits similar patterns to EDCDF but produces a more homogenized temperature field with less defined circulation patterns. ConvLSTM reproduces the general gradient structure but fails to capture finer-scale patterns along the eastern boundary.

**Pre-monsoon** The pre-monsoon season under SSP1 shows a progressive warming pattern across the Bay of Bengal. In March, the initial warming phase begins with temperatures increasing, particularly in the eastern bay. ORAS5 shows that by April, temperatures reach approximately 30$^oC$ in the central and eastern regions, showing the development of the warm feature during this season. By May, which represents peak pre-monsoon conditions, temperatures exceed 31$^oC$ in parts of the central and northern bay. Throughout this period, the CNRM-CM6 model consistently underestimates temperatures by 1-2$^oC$ and fails to accurately capture the formation and spatial

2

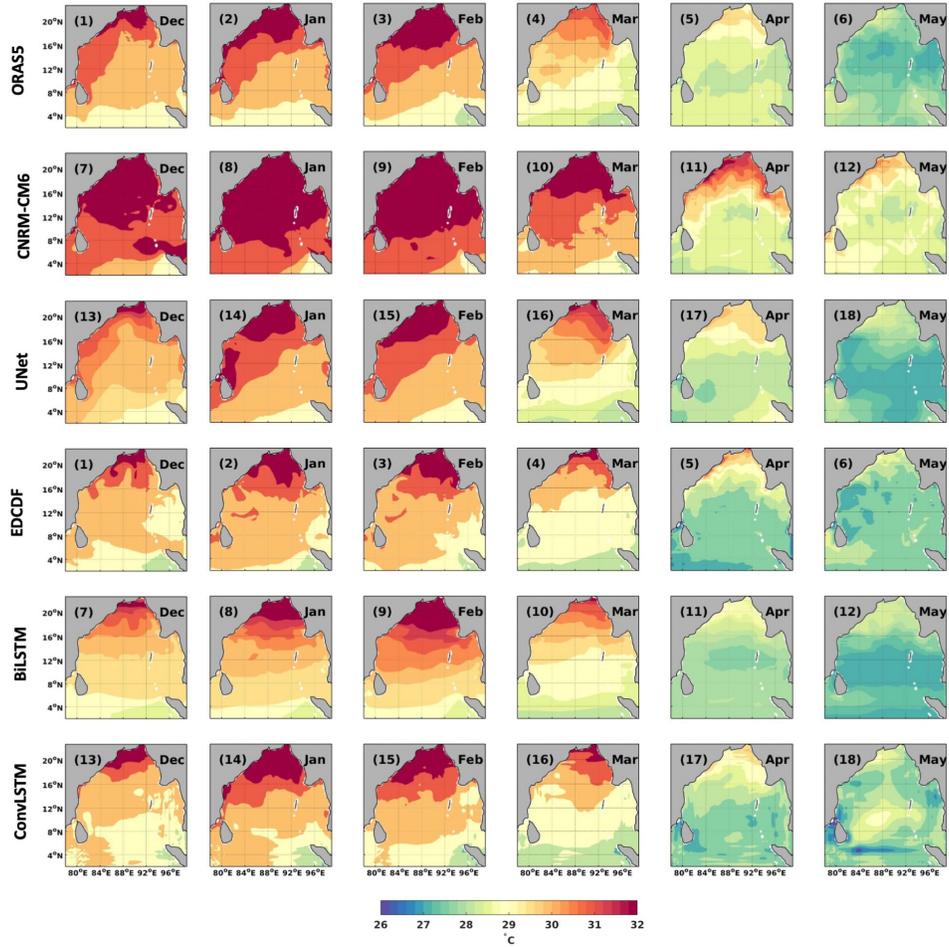

Figure S2: Monthly sea surface temperature (SST) in the BoB during 2021 Dec to May, comparing ORAS5 reanalysis data, raw CNRM-CM6 SSP1-2.6 projections (CNRM-CM6), UNet-corrected SST (UNet), EDCDF corrected SST (EDCDF), BiLSTM corrected SST (BiLSTM), ConvLSTM corrected SST (ConvLSTM).

3

extent of this feature. UNet corrections significantly improve these representations, effectively capturing both the spatial patterns and intensity of warming, particularly in the critical central and eastern regions. The UNet approach successfully reproduces the establishment of the warm pool and its evolution through the pre-monsoon months. EDCDF shows improved performance compared to the winter period but still displays slight temperature overestimation in the northern and western regions during March. BiLSTM provides a more uniform temperature field across the bay but fails to represent mesoscale features in the eastern region. ConvLSTM demonstrates the least effective performance among the correction techniques, failing to adequately capture the essential warming patterns and circulation features in the central bay.

**Monsoon** The monsoon period under SSP1 shows distinctive SST patterns closely associated with monsoon circulation dynamics. June marks the onset of monsoon conditions, with ORAS5 showing temperatures of 29-30$^o$C across most of the bay, with slightly cooler waters in regions directly affected by monsoon winds and currents. July and August display established monsoon patterns, with the Summer Monsoon Current (SMC) evident south of Sri Lanka and the Western Boundary Current flowing northward along the Indian coast. These circulation features create characteristic temperature signatures clearly visible in the ORAS5 data. CNRM-CM6 simulations show a cold bias throughout this period, particularly in the central bay, and fail to accurately represent the thermal signature of the SMC. UNet corrections substantially improve these representations, closely matching ORAS5's depiction of basin-wide patterns and localized features associated with monsoon circulation. Under SSP1, the UNet corrections maintain slightly cooler temperatures in regions affected by upwelling associated with the monsoon circulation, accurately representing these physical processes. EDCDF shows slightly cooler temperatures than observed in the central bay during this period. BiLSTM captures the general spatial patterns but produces a more homogenized structure with less defined circulation features. ConvLSTM generates temperature patterns similar to raw CNRM-CM6 output in August and September, particularly along the western boundary, failing to capture the influence of the East India Coastal Current (EICC).

**Post-monsoon** The post-monsoon transition under SSP1 shows a characteristic progression toward winter conditions. October maintains relatively warm temperatures, with ORAS5 showing values around 29-30$^o$C across much of the bay. November displays the beginning of the winter cooling

4

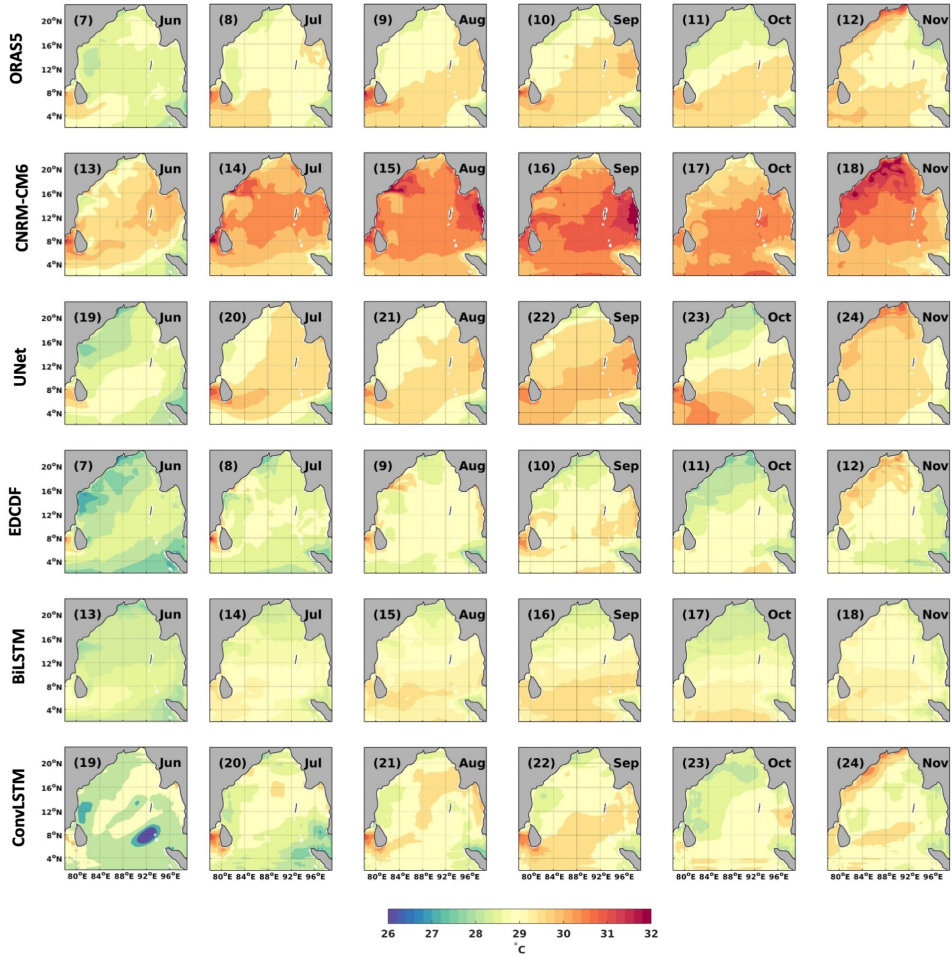

Figure S3: Monthly sea surface temperature (SST) in the BoB during 2021 June to Nov, comparing ORAS5 reanalysis data, raw CNRM-CM6 SSP1-2.6 projections (CNRM-CM6), UNet-corrected SST (UNet), EDCDF corrected SST (EDCDF), BiLSTM corrected SST (BiLSTM), ConvLSTM corrected SST (ConvLSTM).

5

pattern with decreasing temperatures in the northern regions and establishing a north-south gradient. During this period, the East India Coastal Current (EICC) begins its seasonal reversal to southward flow, influencing temperature patterns along the western boundary. CNRM-CM6 exhibits a significant cold bias during this transitional period, particularly in November, where temperatures in the northern bay are underestimated by more than $2^oC$. The UNet corrections demonstrate similar patterns of the ORAS5, accurately capturing both the spatial distribution of temperatures and the onset of winter cooling, and also representing the thermal influence of the EICC along the western boundary. EDCDF shows excessive cooling in the northern bay during November. BiLSTM captures the general transition pattern but fails to represent the finer circulation features, particularly along the western boundary. ConvLSTM fails to accurately capture the development of the north-south gradient in November and misses the thermal signature of the EICC.

## S3 Analysis of UNet Corrected CNRM-CM6 SSP2-4.5 SST Projections in 2021

**Winter** Winter months reveal significant biases in CNRM-CM6 model output compared to ORAS5 reanalysis, with difference plots showing a pronounced cold bias (negative values) of -1- $2^oC$ throughout the northern and central Bay of Bengal. The CNRM-CM6 bias is most severe in January, extending across nearly the entire basin and exceeding $-2^oC$ in the northern regions. Among correction methods, UNet demonstrates the most effective bias reduction, with difference values predominantly within $\pm 0.5^oC$ across most of the basin, though some localized areas in the northern bay still exhibit slight warm bias (0.5- $1^oC$) in December and cold bias in the western regions. EDCDF correction displays an inconsistent performance, with December and January showing a mosaic pattern of positive and negative biases, particularly a warm bias (0.5- $1^oC$) in the northeastern bay and a cold bias (-0.5 to $-1^oC$) in the southwestern regions. BiLSTM correction exhibits a general warm bias (0.5- $1^oC$) in the northern bay during December-January while maintaining near-zero differences in southern regions, indicating overcompensation for the model's cold bias. ConvLSTM correction shows the least consistent performance with alternating patches of warm and cold biases throughout the basin, particularly struggling with the western boundary region where bias values fluctuate between $-1^oC$ and $1^oC$, suggesting difficulty in capturing winter circulation patterns.

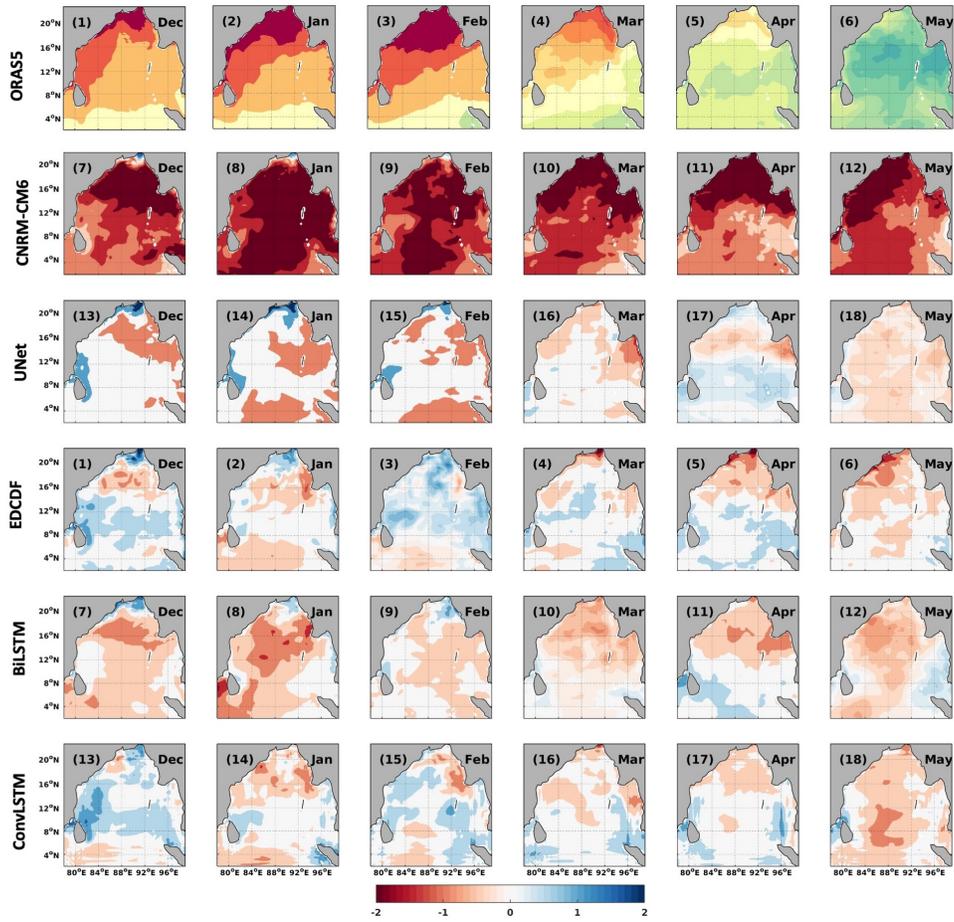

Figure S4: Monthly sea surface temperature (SST) (SSP projection - ORAS5) in the BoB during 2021 Dec to May, ORAS5 reanalysis, difference between ORAS5 with raw CNRM-CM6 SSP2-4.5 (CNRM-CM6), UNet-corrected USSP2-4.5 SST (UNet), EDCDF corrected SST (EDCDF), BiLSTM corrected SST (BiLSTM), ConvLSTM corrected SST (ConvLSTM).

7

**Pre-monsoon** The pre-monsoon transition period difference plots reveal changing bias patterns as the basin warms. CNRM-CM6 exhibits a persistent cold bias throughout March-May, though decreasing in spatial extent compared to winter months, with maximum negative values (-1 to -2$^o$C) concentrated in the central and eastern regions where the warm pool develops. In March, the model cold bias is most widespread, while by May it becomes more localized in the central bay. UNet correction demonstrates excellent performance during this period, with difference values predominantly within ±0.5$^o$C across the basin, though slight cold bias (-0.5$^o$C) appears in the southeastern regions in April-May. EDCDF correction shows mixed performance with March exhibiting warm bias (0.5-1$^o$C) in the northern bay and cold bias (-0.5 to -1$^o$C) in the southern regions, while April and May show improved patterns but still struggle with capturing the warm pool boundary accurately. BiLSTM correction maintains relatively small differences in the central bay but shows inconsistent performance along the boundaries, with positive biases (0.5-1$^o$C) in the northern regions and negative biases (-0.5 to -1$^o$C) in the eastern bay during April-May. ConvLSTM correction exhibits the poorest performance among all methods during pre-monsoon, with significant spatial inhomogeneity in bias patterns and persistent negative differences (-0.5 to -1.5$^o$C) in the southeastern bay during May, failing to capture the critical warm pool development.

**monsoon** Monsoon months display complex bias patterns associated with seasonal circulation features. CNRM-CM6 exhibits its most severe bias during this period, with difference plots showing intense negative values (-1 to -2$^o$C) throughout much of the basin, particularly in July and August when the Summer Monsoon Current (SMC) is most active. The model's cold bias appears most pronounced in the central and western regions influenced by monsoon currents, indicating fundamental issues in representing monsoon circulation. UNet correction achieves remarkable improvement during this challenging period, with difference values predominantly within ±0.5$^o$C across most of the basin, though some residual warm bias (0.5$^o$C) appears in the northwestern bay in June and slight cold bias (-0.5$^o$C) in the southeastern region in August. EDCDF correction shows systematic issues during monsoon months, with persistent cold bias (-0.5 to -1.5$^o$C) in the southern regions across June-September and inconsistent performance along the western boundary where monsoon currents are critical. BiLSTM correction maintains moderate performance with relatively small differences in the central bay but struggles with the spatial complexity of monsoon circulation, showing alternating bands of positive and negative biases along the western boundary.

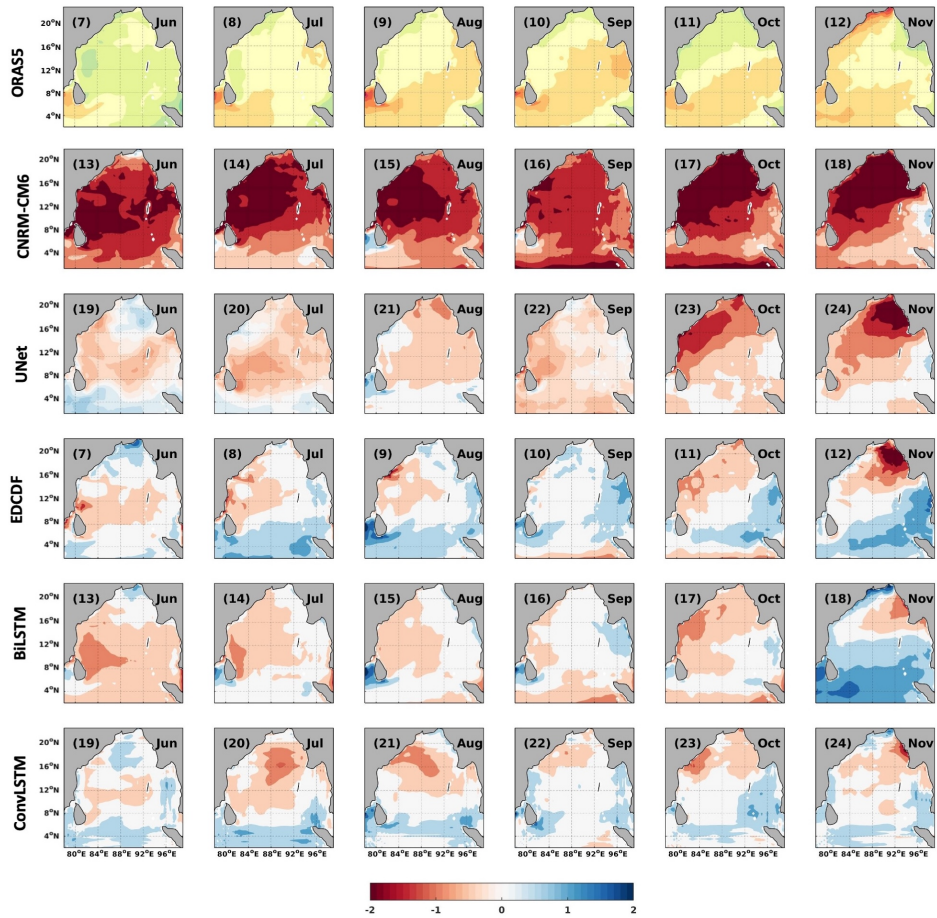

Figure S5: Monthly sea surface temperature (SST) (SSP projection - ORAS5) in the BoB during 2021 June to Nov, ORAS5 reanalysis, Difference between ORAS5 with raw CNRM-CM6 SSP2-4.5 (CNRM-CM6), UNet-corrected USSP2-4.5 SST (UNet), EDCDF corrected SST (EDCDF), BiLSTM corrected SST (BiLSTM), ConvLSTM corrected SST (ConvLSTM).

9

ConvLSTM correction demonstrates particularly poor performance during July-August, with significant patches of negative bias (-1 to -1.5$^o$C) in the southeastern bay and positive bias (0.5-1$^o$C) in the northwestern regions, failing to capture the thermal signatures associated with the SMC and Western Boundary Current.

**Post-monsoon** The post-monsoon transition reveals the most pronounced model biases and challenging correction scenarios of all seasons. CNRM-CM6 exhibits extreme cold bias during this period, with difference plots showing intense negative values extending across nearly the entire basin, reaching -2$^o$C or below in the northern and central regions during November when the winter cooling pattern begins to establish. This suggests fundamental issues in representing the post-monsoon transition and EICC reversal. UNet correction maintains impressive performance even during this difficult period, with difference values predominantly within ±0.5$^o$C in October, though November shows more substantial residual biases with warm differences (0.5-1$^o$C) in the northern bay and cold differences (-0.5 to -1$^o$C) along portions of the western boundary. EDCDF correction displays significant issues during post-monsoon, with October showing a complex mosaic of biases and November exhibiting severe cold bias (-1 to -2$^o$C) in the northern regions, suggesting inability to capture the developing north-south gradient. BiLSTM correction shows moderate performance in October but deteriorates significantly in November, with intense negative bias (-1 to -2$^o$C) in the southern half of the bay, completely missing the EICC reversal thermal signature. ConvLSTM correction demonstrates the least effective performance among all methods during post-monsoon, with persistent negative bias (-0.5 to -1.5$^o$C) throughout most of the basin in both months, with slight positive differences in scattered, physically inconsistent patches, indicating fundamental limitations in representing the complex post-monsoon dynamics and the onset of winter cooling.

# S4 Analysis of UNet Corrected CNRM-CM6 SSP3-7.0 SST Projections in 2021

Figure S6 and S7 displays the monthly SST for 2021 from reanalysis (ORAS5), raw CNRM-CM6 SSP3-7.0 projections (CNRM-CM6), UNet-corrected projections (UNet), EDCDF corrected SST (EDCDF), BiLSTM corrected SST (BiLSTM), and ConvLSTM corrected SST (ConvLSTM) in the Bay of Bengal.

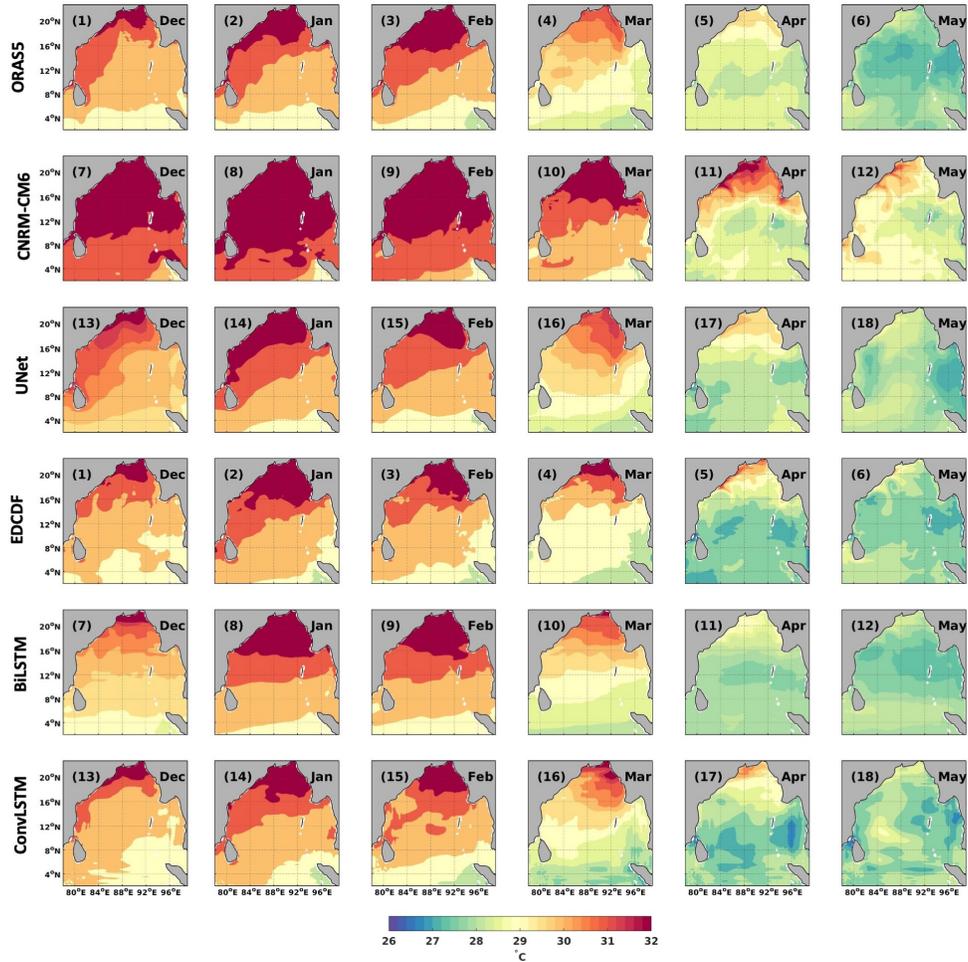

Figure S6: Monthly sea surface temperature (SST) in the BoB during 2021 Dec to May, comparing ORAS5 reanalysis data, raw CNRM-CM6 SSP3-7.0 projections (CNRM-CM6), UNet-corrected SST (UNet), EDCDF corrected SST (EDCDF), BiLSTM corrected SST (BiLSTM), ConvLSTM corrected SST (ConvLSTM).

11

**Winter** Under the SSP3 scenario, winter months in the Bay of Bengal demonstrate a more pronounced temperature gradient with intensified cooling in the northern regions. ORAS5 observations continue to show temperatures of 26- 27ºC in the northern bay with a gradual transition to 28- 29ºC in the southern regions, but the CNRM-CM6 model exhibits a more severe cold bias under this higher-emissions pathway. The spatial extent of waters below 26ºC is significantly exaggerated, covering most of the northern and central bay. Among correction techniques, UNet again demonstrates superior performance, effectively mitigating this enhanced cold bias and restoring temperature patterns close to the ORAS5 reanalysis. The UNet corrections successfully capture the appropriate north-south gradient and temperature boundaries, though some residual cold bias remains in the northwestern bay during January. EDCDF shows mixed performance with a tendency toward overcorrection in the northern regions, producing temperatures exceeding ORAS5 by 0.5- 1ºC in December and January. BiLSTM performs adequately in maintaining the general gradient structure but produces a more homogenized temperature field with reduced spatial complexity. ConvLSTM struggles to correct the enhanced cold bias of the CNRM-CM6 under SSP3, particularly along the eastern boundary, where mesoscale features are poorly resolved.

**Pre-monsoon** The pre-monsoon season under SSP3 exhibits intensified warming patterns compared to other scenarios, reflecting the higher-emissions pathway. March shows the initial warming phase, but the progression is more rapid than in other scenarios. By April, temperatures in the ORAS5 reanalysis reach 30- 31ºC in the central and eastern regions, with the warm pool development more pronounced and spatially extensive. May represents peak pre-monsoon conditions with temperatures exceeding 31.5ºC across much of the central and northern bay. The CNRM-CM6 model shows a persistent cold bias throughout this period. UNet corrections substantially improve these representations, accurately capturing both the intensified warming and the expanded spatial extent of the warm pool. The UNet approach successfully reproduces the accelerated warming trajectory under SSP3, particularly in the critical central and eastern regions. EDCDF shows moderate performance during this period but struggles with the spatial complexity of the warming pattern, particularly in April when the warm pool begins its rapid expansion. BiLSTM produces a more uniform temperature field that fails to capture the intensified warming gradients under SSP3. ConvLSTM demonstrates the poorest performance among correction techniques, unable to adequately represent the enhanced warming patterns and circulation fea-

tures in the central and eastern bay that characterize the SSP3 pre-monsoon season.

**Monsoon** The monsoon period under SSP3 shows altered temperature patterns associated with intensified monsoon circulation dynamics. June marks the monsoon onset with ORAS5 showing generally elevated temperatures of 29-30.5ºC across most of the bay. July and August display intensified monsoon patterns, with the Summer Monsoon Current (SMC) showing a stronger thermal signature south of Sri Lanka and enhanced upwelling along the western boundary. CNRM-CM6 exhibits an augmented cold bias throughout this period under the SSP3 scenario, particularly in the central bay, and critically fails to represent the intensified SMC's thermal signature. UNet corrections substantially improve these representations, successfully capturing both the basin-wide warming and the localized cooling features associated with the strengthened monsoon circulation under SSP3. EDCDF struggles with the spatial complexity of these patterns, showing inconsistent performance with both over- and under-correction in different regions of the bay. BiLSTM captures the general basin-wide patterns but produces a more homogenized structure that fails to represent the enhanced circulation features under SSP3. ConvLSTM generates temperature patterns that retain much of the cold bias of the raw CNRM-CM6 output, particularly along the western boundary, and fails to adequately capture the intensified influence of SMC and the East India Coastal Current (EICC).

**Post-monsoon** The post-monsoon transition under SSP3 displays more rapid cooling and enhanced temperature gradients compared to other scenarios. October maintains elevated temperatures with ORAS5 showing values around 29.5-30.5ºC across much of the bay, but November exhibits accelerated cooling with a more pronounced north-south gradient. The East India Coastal Current (EICC) shows a stronger signal during its seasonal reversal to southward flow, creating more distinct temperature patterns along the western boundary. CNRM-CM6 exhibits its most significant cold bias during this transitional period under SSP3, particularly in November, where temperatures in the northern bay are underestimated, and the spatial extent of the cold bias extends further southward. The UNet corrections demonstrate exceptional skill in reproducing these complex patterns, accurately capturing both the intensified cooling trajectory and the enhanced spatial gradients while properly representing the stronger thermal signature of the EICC along the western boundary. EDCDF shows excessive cooling in the northern bay during November and fails to capture the enhanced gradient structure. BiL-

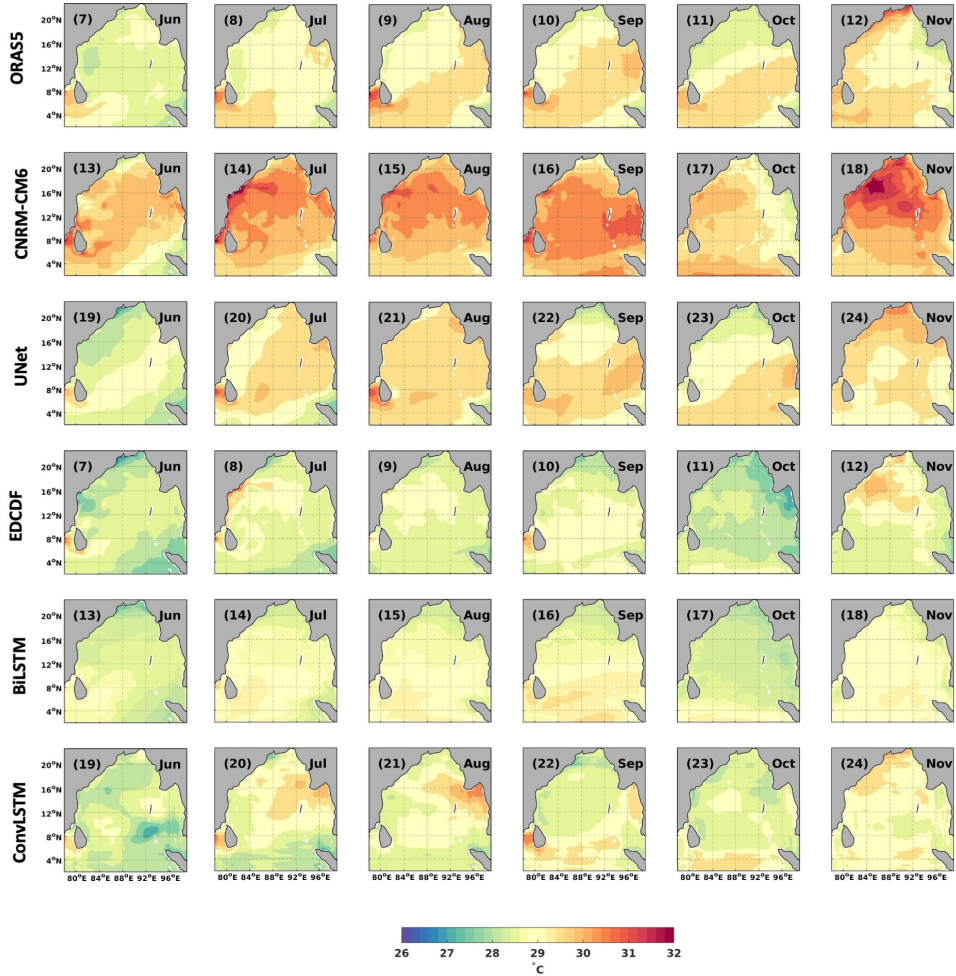

Figure S7: Monthly sea surface temperature (SST) in the BoB during 2021 June to Nov, comparing ORAS5 reanalysis data, raw CNRM-CM6 SSP3-7.0 projections (CNRM-CM6), UNet-corrected SST (UNet), EDCDF corrected SST (EDCDF), BiLSTM corrected SST (BiLSTM), ConvLSTM corrected SST (ConvLSTM).

STM captures the general cooling pattern but significantly underrepresents the intensified spatial complexity and circulation features. ConvLSTM fails to adequately correct the enhanced cold bias of CNRM-CM6, particularly in November, and misses the strengthened thermal signature of the EICC that characterizes the post-monsoon transition.

## S5   Analysis of UNet Corrected CNRM-CM6 SSP5-8.5 SST Projections in 2021

Figure S8 and S9 displays the monthly SST for 2021 from reanalysis (ORAS5), raw CNRM-CM6 SSP5-8.5 projections (CNRM-CM6), UNet-corrected projections (UNet), EDCDF corrected SST (EDCDF), BiLSTM corrected SST (BiLSTM), and ConvLSTM corrected SST (ConvLSTM) in the Bay of Bengal.

**Winter**   The winter period under the SSP5 scenario demonstrates the most extreme temperature patterns among all pathways, reflecting the fossil-fuel intensive development trajectory. While ORAS5 reanalysis shows the characteristic north-south temperature gradient with values of 26- 27ºC in the northern bay increasing to 28- 29ºC in southern regions, the CNRM-CM6 model exhibits its most severe cold bias under SSP5. Among correction techniques, UNet reduces the cold bias and restores temperature patterns that resemble ORAS5 reanalysis. EDCDF struggles with the magnitude of corrections required under SSP5, showing both over-correction in some areas of the northern bay and under-correction in others, particularly along the eastern boundary. BiLSTM produces a more spatially homogeneous correction that captures the general pattern but fails to represent the complex thermal structures that characterize the bay during winter. ConvLSTM shows its most significant limitations under SSP5 conditions, failing to adequately address the extreme cold bias in the central and northern regions and producing spurious thermal features along the eastern and western boundaries.

**Pre-monsoon**   The pre-monsoon season under SSP5 exhibits the most dramatic warming progression and highest temperature extremes among all scenarios. March initiates a rapid warming phase, with temperatures increasing more quickly than in other pathways. By April, ORAS5 data shows temperatures exceeding 31ºC in extensive areas of the central and eastern bay, with the warm pool development both more intense and spatially extensive than in other scenarios. May represents peak pre-monsoon conditions

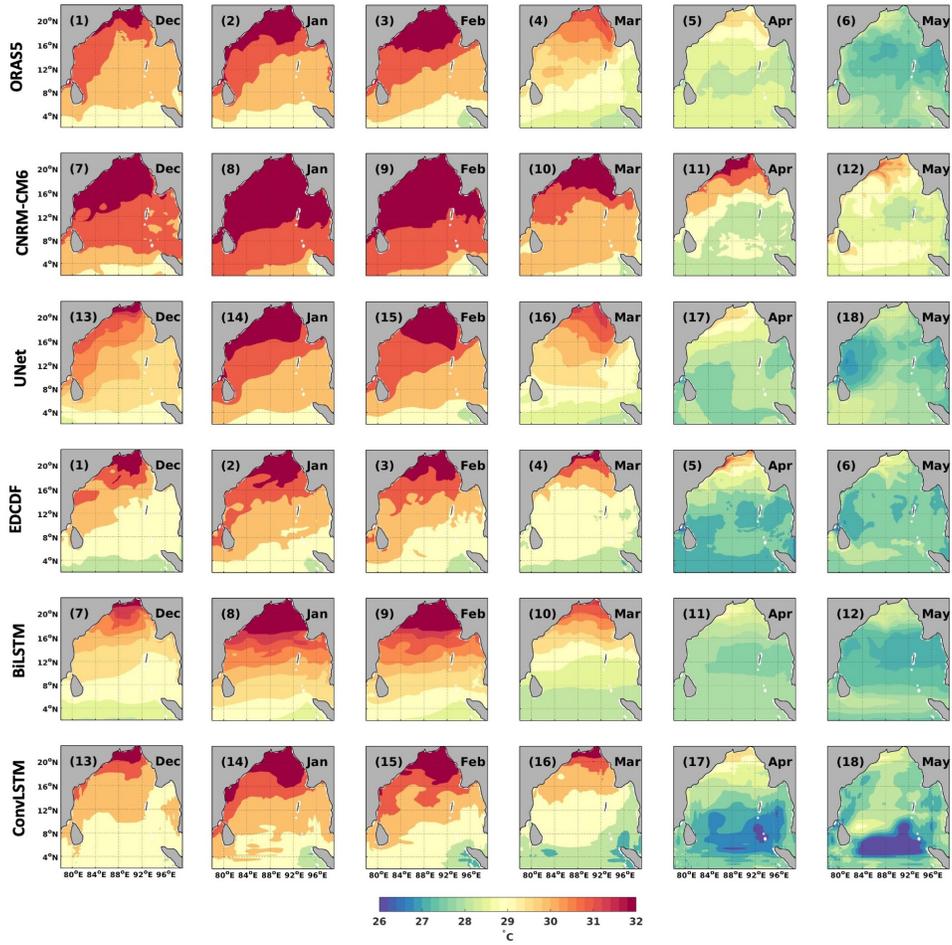

Figure S8: Monthly sea surface temperature (SST) in the BoB during 2021 Dec to May, comparing ORAS5 reanalysis data, raw CNRM-CM6 SSP5-8.5 projections (CNRM-CM6), UNet-corrected SST (UNet), EDCDF corrected SST (EDCDF), BiLSTM corrected SST (BiLSTM), ConvLSTM corrected SST (ConvLSTM).

16

with temperatures reaching up to 32$^o$C in parts of the northern and central bay, creating unprecedented thermal conditions. CNRM-CM6 shows a severe cold bias throughout this period, with temperatures underestimated by 2-3$^o$C across much of the basin, completely failing to capture the enhanced warm pool development characteristic of SSP5. This bias fundamentally misrepresents the extreme thermal conditions that would characterize this fossil-fuel-intensive pathway. UNet corrections address these extreme biases, successfully reproducing both the accelerated warming trajectory and the expanded warm pool, particularly in the critical central and eastern regions. The UNet approach effectively captures the unprecedented thermal extremes projected under SSP5. EDCDF shows inconsistent performance, with adequate corrections in some regions but significant errors in others, particularly in May when the thermal extremes are most pronounced. BiLSTM produces a more uniform temperature field that fails to capture the enhanced spatial complexity and extreme values characteristic of SSP5. ConvLSTM exhibits systematic deficiencies in representing the extreme thermal conditions, producing unrealistic cool anomalies in the southeastern bay during April and May that have no physical basis in either the observations or raw CNRM-CM6 SSP5 output.

**Monsoon** The monsoon period under SSP5 exhibits the most dramatically altered temperature patterns, reflecting the projected intensification of monsoon dynamics under this high-emissions pathway. June shows an earlier and more intense onset of monsoon conditions, with ORAS5 displaying a complex pattern of elevated baseline temperatures (30-31$^o$C) interrupted by more pronounced cooling in regions directly affected by the significantly strengthened monsoon circulation. July and August demonstrate intensified monsoon patterns with the Summer Monsoon Current (SMC) showing an enhanced thermal signature south of Sri Lanka and more extreme upwelling along the western boundary. September shows a delayed monsoon withdrawal with sustained thermal extremes across the bay. CNRM-CM6 exhibits its most significant biases under SSP5, with temperatures underestimated by 2-3$^o$C across much of the basin while failing to represent the intensified circulation features characteristic of this scenario. UNet corrections show remarkable skill in reproducing these complex patterns, capturing both the elevated baseline temperatures and the enhanced cooling signatures associated with intensified monsoon circulation. EDCDF shows highly variable performance, with adequate corrections in some regions but significant errors in others, particularly failing to capture the enhanced mesoscale features associated with intensified circulation. BiLSTM produces overly homogenized

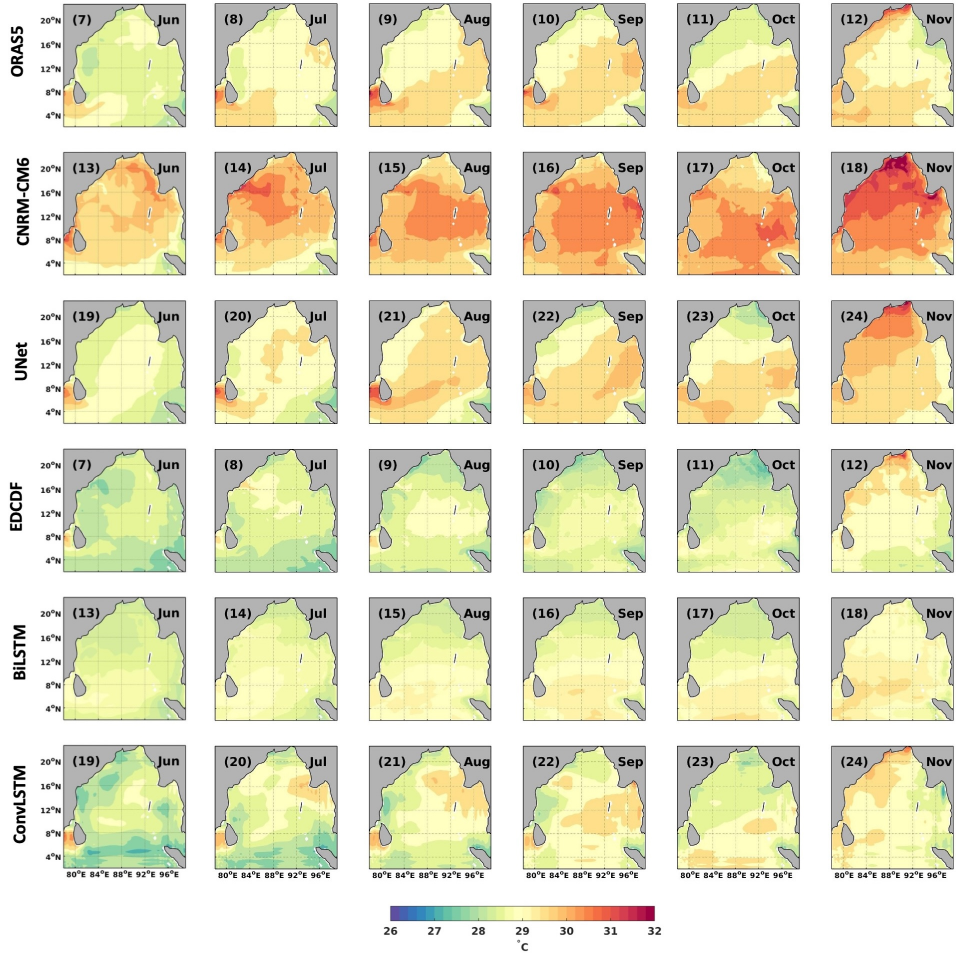

Figure S9: Monthly sea surface temperature (SST) in the BoB during 2021 June to Nov, comparing ORAS5 reanalysis data, raw CNRM-CM6 SSP5-8.5 projections (CNRM-CM6), UNet-corrected SST (UNet), EDCDF corrected SST (EDCDF), BiLSTM corrected SST (BiLSTM), ConvLSTM corrected SST (ConvLSTM).

temperature fields that mask the critical circulation features characteristic of monsoon conditions under SSP5. ConvLSTM demonstrates systematic deficiencies in representing the complex spatial patterns, particularly along the western boundary, where the intensified East India Coastal Current (EICC) creates distinctive thermal signatures that this correction method fails to reproduce.

**Post-monsoon** The post-monsoon transition under SSP5 displays the most extreme cooling trajectory and temperature gradients among all scenarios. October maintains significantly elevated temperatures, with ORAS5 showing values exceeding 30- 31ºC across much of the bay, but November exhibits an accelerated cooling pattern with a dramatically enhanced north-south gradient establishing earlier and more intensely than in other scenarios. The East India Coastal Current (EICC) displays its strongest signal during its seasonal reversal, creating highly distinctive temperature patterns along the western boundary. CNRM-CM6 exhibits catastrophic biases during this period under the SSP5 scenario. UNet corrections demonstrate exceptional skill in addressing these extreme biases, successfully reproducing both the delayed cooling trajectory and the enhanced spatial gradients while accurately representing the intensified thermal signature of the EICC along the western boundary. Under SSP5, the UNet approach effectively captures the unprecedented temperature extremes and their spatial distribution that characterize this fossil-fuel intensive pathway. EDCDF shows significant limitations in addressing the magnitude of biases present, particularly in November, with over- and under-correction in different regions. BiLSTM captures the general cooling pattern but substantially underrepresents the spatial complexity and extreme gradients characteristic of SSP5. ConvLSTM fails dramatically in representing the post-monsoon transition under this scenario, producing unrealistic temperature patterns, particularly in the northern bay, where the cooling dynamics are most pronounced.



\end{document}